\documentclass[11pt]{article}
\pdfoutput=1
\usepackage{jheppub}
\usepackage{amsmath}
\usepackage{bm}
\usepackage{slashed}
\usepackage{graphicx}

\newcommand{\tr}{\mathrm{tr}\,}

\newlength{\dummysp}
\newcommand{\beq}{\begin{eqnarray}}
\newcommand{\eeq}{\end{eqnarray}}

\newcommand{\e}{{\epsilon}}

\newcommand{\gappeq}{\mathrel{\rlap {\raise.5ex\hbox{$>$}}
{\lower.5ex\hbox{$\sim$}}}}
\newcommand{\lappeq}{\mathrel{\rlap{\raise.5ex\hbox{$<$}}
{\lower.5ex\hbox{$\sim$}}}}

\newcommand{\ben}{\begin{enumerate}}
\newcommand{\een}{\end{enumerate}}

\newcommand{\bit}{\begin{itemize}}
\newcommand{\eit}{\end{itemize}}

\def\[{\left [}
\def\]{\right ]}
\def\({\left (}
\def\){\right )}
\def\R{{\mathbb R}}
\def\S{{\mathbb S}}
\def\Z{{\mathbb Z}}

\title{Domain walls in high-$\bm T$ $\bm{SU(N)}$ super Yang-Mills theory and QCD(adj)}
   
 \author[a]{Mohamed M. Anber,}\author[b]{Erich Poppitz} 
  
\affiliation[a]{Department of Physics, Lewis $\&$ Clark College, Portland, OR 97219, USA}
\affiliation[b]{Department of Physics,   University of Toronto, 
Toronto, ON M5S 1A7, Canada}
\emailAdd{manber@lclark.edu}\emailAdd{poppitz@physics.utoronto.ca}    

\abstract{

{\flushleft{W}}e study the domain walls in hot $4$-D $SU(N)$ super Yang-Mills theory and QCD(adj), with $n_f$ Weyl flavors. We find that the $k$-wall worldvolume theory is   $2$-D QCD with gauge group $SU(N-k)\times SU(k) \times U(1)$ and Dirac fermions charged under  $U(1)$ and transforming in the bi-fundamental representation of the nonabelian factors. We show that the DW theory has a $1$-form $\mathbb Z_{N}^{(1)}$ center symmetry and a $0$-form $\mathbb Z_{2Nn_f}^{d\chi}$ discrete chiral symmetry, with a mixed 't Hooft anomaly consistent with bulk/wall anomaly inflow. We argue that $\mathbb Z_{N}^{(1)}$ is broken on the wall, and hence, Wilson loops obey the perimeter law. The breaking of the worldvolume center symmetry implies that bulk $p$-strings can end on the wall, a phenomenon first discovered using string-theoretic constructions. We invoke $2$-D bosonization  and gauged Wess-Zumino-Witten models to  suggest that $\mathbb Z_{2Nn_f}^{d\chi}$ is also broken in the IR, which implies that the $0$-form/$1$-form mixed 't Hooft anomaly in the gapped $k$-wall theory is saturated by a topological quantum field theory. We also find interesting parallels   between the physics of high-temperature domain walls studied here and domain walls between chiral symmetry breaking vacua in the zero temperature phase of the theory (studied earlier in the semiclassically calculable small spatial circle regime), arising  from the similar mode of saturation of the relevant 't Hooft anomalies.}

\begin{document}

\maketitle

\flushbottom

\section{Introduction}

Domain walls (DW) are ubiquitous in field theory as they appear in many natural phenomena, ranging from condensed matter physics to cosmology, due to the spontaneous breaking of global symmetries. Among the plethora of field theories, $SU(N)$ Yang-Mills (YM) theory and its ${\cal N}=1$ supersymmetric generalization (SYM) stand out as they play an important role in the Standard Model and its extensions. These theories are invariant under a $\mathbb Z_N^{(1)}$ discrete one-form global symmetry known as center symmetry.\footnote{We use a superscript  $\mathbb Z_N^{(1)}$to distinguish one-form symmetries from ordinary zero-form symmetries such as the discrete chiral $\mathbb Z_{2N}$.  For an introduction to center symmetry and to its relevance as an order parameter for confinement, from lattice and continuum perspectives, see \cite{Greensite:2011zz}, \cite{Gaiotto:2014kfa}.} 

At temperature  $T$ below the deconfinement temperature $T_c$ (of order the strong scale $\Lambda$) the expectation value of the Polyakov loop $P$---which is charged under $\mathbb Z^{(1)}_N$---vanishes, signalling that the theory is in the confined phase. At temperature greater than $T_c$ the theory deconfines and the center symmetry breaks spontaneously, giving rise to DW that interpolate between $N$ distinct vacua, which are distinguished by the expectation value of the Polyakov's loop: $\langle P \rangle=Ne^{-i\frac{2\pi k}{N}}$, and $k=0,1,2,...,N-1$. 

These DW are closely related to center vortices, which are thought to be responsible for disordering the vacuum and giving rise to confinement as the temperature is decreased below $T_c$, see \cite{Greensite:2011zz} for an introduction and review. Therefore, one hopes that a close examination of the DW will shed light into the role of center vortices in the strong dynamics. Fortunately enough, DW are amenable to perturbative analysis at $T\gg T_c$, which makes them excellent objects to study compared to their low-temperature counterparts, the center vortices.\footnote{These do not yield to controlled analytic studies, but require lattice simulations or model assumptions \cite{Greensite:2011zz}.}

 Despite the fact that DW in Yang-Mills theory  are well studied in the literature, in the perturbative regime \cite{Bhattacharya:1990hk, Bhattacharya:1992qb, Smilga:1993vb,KorthalsAltes:1999xb,KorthalsAltes:2000gs,Armoni:2008yp,KorthalsAltes:2009dp,Armoni:2010ny,Draper:2018mpj,Ritz:2018mce}, on the lattice \cite{deForcrand:2004jt,Bursa:2005yv}, via holography \cite{Aharony:1998qu,Armoni:2008yp,Armoni:2010ny,Argurio:2018uup}, or with an emphasis on DW in  supersymmetric Yang-Mills theory \cite{Acharya:2001dz, 
Ritz:2002fm, Tong:2003ik, Ritz:2004mp, Hanany:2005bq, Bashmakov:2018ghn}, the DW worldvolume theory and the interplay between the bulk and DW physics remain, in many cases, a largely unexplored territory.\footnote{Many of these studies  focused on the $k$-wall tension, argued to exhibit Casimir scaling as in (\ref{k-wall action}).}
A renewed impetus for such studies is provided by the recent realization that  DW must have  rich worldvolume dynamics, required by newly discovered anomalies \cite{Gaiotto:2014kfa,Gaiotto:2017yup,Gaiotto:2017tne}, as   discussed further below.  

 In recent work \cite{Anber:2018jdf}, we studied the DW in high-T $SU(2)$ ${\cal N}=1$ super-YM theory to find that the two-dimensional  ($2$-D) worldvolume theory is given by the axial version of the charge-$2$ Schwinger model.   This theory was shown to  have a broken $\mathbb Z_{4}^{d\chi}$ discrete chiral symmetry\footnote{The ``$d \chi$" superscript is a reminder of the nature of the discrete symmetry.}  and a broken $\mathbb Z^{(1)}_{2}$ center symmetry. The broken  chiral symmetry on the wall implies that the fermion bilinear condensate on the wall should be nonzero in the high-$T$, chirally restored and deconfined  phase of the bulk. The broken center symmetry on the wall implies a perimeter law for a fundamental Wilson loop. This  behavior on the $2$-D   worldvolume mirrors many properties of the strongly coupled $4$-D   low temperature theory, inferred from its $M$-theory embedding \cite{Witten:1997ep} or from  weakly coupled $\R^3 \times \S^1$ compactifications \cite{Anber:2015kea}.\footnote{These calculable compactifications are in many cases continuously connected to the $\R^4$ strongly coupled theory, see \cite{Dunne:2016nmc} for a recent review and references.}

Motivated by the rich structure of the DW in the $SU(2)$ case, in this paper we generalize our study to the $k$-walls in $SU(N)$ ${\cal N}=1$ super-YM theory, examine their worldvolume theory and the fate of the various discrete symmetries. The generalization   to  $N>2$ presents various technical challenges  addressed  in the Appendices. We also fill in many details left out in \cite{Anber:2018jdf}. 
Many of the results we find also apply to YM theory with higher supersymmetry as well as to their non-supersymmetric versions with multiple adjoint Weyl fermions, QCD(adj). 

An important tool in our study is the use of the 't Hooft anomaly matching conditions \cite{tHooft:1980xss,Frishman:1980dq,Coleman:1982yg,Csaki:1997aw}. Given a global symmetry $G$ of a quantum field theory, the gauging of this symmetry may be obstructed due to the existence of an anomaly.  The obstruction is renormalization group invariant and can be used to set constraints on the IR spectrum of the theory, which are  particularly useful in asymptotically free theories. Most relevant to our study is the fact that a new type of 't Hooft anomaly was recently discovered in \cite{Gaiotto:2014kfa,Gaiotto:2017yup,Gaiotto:2017tne}. This is a mixed  anomaly between two discrete global symmetries such that one becomes anomalous as we gauge the other. One of the two is a $0$-form  symmetry, which means that it acts on local operators, while the other is a $1$-form symmetry that acts on line operators, e.g., Wilson loops. Anomalies of this new type have been the subject of many recent investigations (for an incomplete list, see~\cite{Komargodski:2017dmc,Komargodski:2017smk,Shimizu:2017asf,Kikuchi:2017pcp,Tanizaki:2017qhf,Tanizaki:2017mtm,Cherman:2017dwt,Sulejmanpasic:2018upi,Tanizaki:2018xto,Benini:2018reh,Aitken:2018kky,Anber:2018tcj,Cordova:2018acb,Tanizaki:2018wtg,Bi:2018xvr,Choi:2018tuh,Yamaguchi:2018xse}).

One of the striking findings in this work is that various $2$-D gauge theories with Dirac fermions,  thought to be just ``toy models"  extensively studied for their similarity with $4$-D QCD (see \cite{Frishman:1992mr,Frishman:2010zz} for reviews) are tied to the full-fledged $4$-D super Yang-Mills theory and its various supersymmetric and nonsupersymmetric extensions   via its DW worldvolume theory.  
In particular, we show that the worldvolume of the $k$-wall is  a $2$-D $U(1)\times SU(N-k)\times SU(k)$ YM theory with Dirac fermions, charged under the $U(1)$ and transforming in the  bi-fundamental representation of $SU(N-k) \times SU(k)$. We show that this theory has an anomaly free $0$-form $\mathbb Z_{2N}^{d\chi}$ discrete  chiral  symmetry, while fundamental Wilson loops transform under a   ${\mathbb Z}_N^{(1)}$ center symmetry. We argue that  fundamental Wilson loops exhibit  a perimeter law, and hence,  the fundamental quarks are deconfined on the wall. The bulk, on the other hand, is a strongly coupled non-Abelian  $3$-D gauge theory which possesses a mass gap and confines.\footnote{This has nothing to do with confinement of  real heavy quarks in the original $3+1$-D theory.} Consequently,  one can turn on $p$-flux tubes in the bulk, sourced by $N$-ality $p$ probe quarks, and examine their behavior as they join the wall.  We argue that these tubes will terminate on the wall as a consequence of the screening of fundamental charges (perimeter law) on the wall: as a $p$-tube joins the wall it will break into representations of $U(1)\times SU(k)\times SU(N-k)$, which are screened by the DW fermions.   That confining strings can end on domain walls was first discovered in the context of M-theory  \cite{Witten:1997ep}, for low-$T$ DW associated with the breaking of the discrete chiral $R$-symmetry of super-YM theory,   and via holography in ${\cal N}=4$ super-YM \cite{Aharony:1998qu}, for the high-$T$ DW studied here. Our study gives the first weakly coupled  high-$T$ field theory dynamical explanation of this phenomenon. 

Previously, a weakly coupled field theory mechanism explaining how confining strings can end on  low-$T$  DW (due to $R$ symmetry breaking) was found in \cite{Anber:2015kea} in the context of $\R^3 \times \S^1$ compactifications. Here, we find that there are many similarities between the properties of DW in the  two small-$\S^1$ cases---small spatial circle vs. high-$T$--- due to the similar ways that  't Hooft anomalies are saturated,  see Section \ref{section:chiralanomaly} for further discussion (as well as Figure \ref{fig:01}). 
Obtaining a better understanding of the microscopic mechanism allowing strings to end on DW and of its relation to anomalies and inflow in more general cases than considered so far (for example, on $\R^4$ \cite{Acharya:2001dz}; see \cite{Dierigl:2014xta} for topological arguments within QFT) is an interesting task for future studies.

 The fate of the discrete chiral symmetry on the $k$-wall is more subtle since the $k$-wall worldvolume theory for $SU(N)$, as opposed to $SU(2)$,  is not exactly solvable. However, arguments based on bosonization and gauged Wess-Zumino-Witten models suggest that $\mathbb Z_{2N}^{d\chi}$  is spontaneously broken to $\mathbb Z_{2}^{d\chi}$ (the latter is part of the Lorentz symmetry)   giving rise to $N$ distinct vacua on the $k$-wall, which are needed to saturate the mixed discrete chiral/center anomaly. This means that the IR theory is ``empty", i.e., it has no massless degrees of freedom, and the 't Hooft anomalies are matched by a topological quantum field theory (TQFT).

 This paper is organized as follows. In Section \ref{Domain walls, anomalies, and inflow} we examine the DW in high-$T$ super Yang-Mills and derive the worldvolume theory (we do not consider the decoupled center of mass degrees of freedom). We then study the discrete symmetries of the $k$-wall worldvolume theory,  and show that it has a $0$-form/$1$-form mixed 't Hooft anomaly, which is also consistent with the bulk/wall anomaly inflow. In Section \ref{Screening and strings ending on walls}, we study the realization of the $\mathbb Z^{(1)}_{N}$ center symmetry on the wall and show that it is broken, hence Wilson lines obey the perimeter law. We then argue that the bulk $p$-string can end on the DW. We study the fate of the discrete chiral symmetry and discuss the IR TQFT saturating the anomaly in Section \ref{section:chiralanomaly}, while in Section \ref{k-walls in QCD(adj)} we comment on the $k$-walls in adjoint QCD. 
 
 Many important technical details are relegated to Appendices. We summarize our group theory conventions in Appendix \ref{appendix:groups} and work out the details of the DW fermion zero-modes in Appendix \ref{appendix:zeromodes}. Results crucial for understanding the anomalies of the chiral and  center symmetry of the $k$-wall theory---the $U(1)$-flux quantization and 't Hooft fluxes---are derived in  Appendix \ref{appendix:quantization}, where we study the properties of the $SU(N-k)\times SU(k)\times U(1)$ bundle\footnote{\label{groupfootnote}More precisely, the $k$-wall worldvolume gauge group is $SU(N-k)\times SU(k) \times U(1)/(\Z_{N-k} \times \Z_k)$.} on the torus, and in Appendix \ref{appendix:thooftprojection} using a projection of  constant flux backgrounds.

\section{Domain walls, anomalies, and inflow}
\label{Domain walls, anomalies, and inflow}

\subsection{Adjoint QCD at high temperature}

We consider $SU(N)$ Yang-Mills theory endowed with $n_f$ adjoint Weyl fermions at finite temperature $T$:
\begin{eqnarray}
S= \frac{1}{g^2}\int_{\mathbb R^3\times \mathbb S_\beta^1}\frac{1}{2}\;\mbox{tr}_F\left( F_{\mu\nu}F_{\mu\nu}\right)+i\mbox{tr}_F\left(\bar \lambda \bar \sigma^\mu D_\mu\lambda\right)\,,
\label{SUN Lagrangian}
\end{eqnarray}
where $\mu,\nu=1,2,3,4$ and the fundamental trace is normalized as $\mbox{tr}_F\left(t^at^b\right)=\delta_{ab}$.  In this normalization the roots have length $\bm \alpha^2=2$.  $\mathbb S_\beta^1$ is the thermal circle, which is taken along the $x_4$-direction and has circumference $\beta=1/T$. The covariant derivative is given by $D_\mu=\partial_\mu -i[A_\mu,\,\,]$ and $\bar \sigma=\left(\bm\sigma, -i\right)$, where $\bm \sigma$ are the spacetime Pauli matrices.  In addition, the fermion field $\lambda$ carries an implicit flavor index. 

At temperatures larger than  $\Lambda$, the strong coupling scale of the theory, many aspects of the theory become amenable to semiclassical treatment owing to asymptotic freedom. In this case we can dimensionally reduce the action (\ref{SUN Lagrangian}) to $3$-D after integrating out a tower of heavy Matsubara excitations of the gauge and fermion fields along $\mathbb S_1^\beta$. To one-loop order, the resulting bosonic part of the action \cite{Gross:1980br} reads 
\begin{eqnarray}
S_{3-D}^{{\rm boson}}=\frac{\beta}{g^2(\beta)}\int_{\mathbb R^3}\left( \frac{1}{2}\mbox{tr}_F\left( F_{ij}F_{ij}\right)+\mbox{tr}_F\left(D_i  A_4\right)^2+g^2V(\bm A_4)+{\cal O}\left(g^4\right) \right) \,,
\label{bosonic action}
\end{eqnarray}
where $i,j=1,2,3$ and $V(\bm A_4)$ is the one-loop effective potential for the Matsubara zero mode of the $x^4$-component of the gauge field written  in terms of its Cartan subalgebra component $\bm A_4$:
\begin{eqnarray}
V(\bm A_4)=\frac{4T^4}{\pi^2}\sum_{\bm \beta^+}\sum_{n=1}\frac{-1+ n_f (-1)^n}{n^4}\cos\left[\frac{n \bm A_4\cdot \bm \beta}{T}\right]\,,
\label{total V}
\end{eqnarray}
and the sum is over all positive roots $\bm \beta_+$ (to not be confused with the inverse temperature $\beta = 1/T$). Our group theory conventions are detailed in Appendix \ref{appendix:groups}. In the rest of this paper we consider SYM (i.e., $n_f=1$), while we discuss $n_f>1$ in Section \ref{k-walls in QCD(adj)}. 
  
 \subsection{Vacua and domain walls}

The potential (\ref{total V}) has $N$ vacua, all with  $SU(N)$  unbroken:
\begin{eqnarray}
\langle \bm A_4\rangle_{(a)} \beta \equiv \bm\Phi_0=2\pi \bm\omega_a , \quad a=0,1,2,...,N-1\,, 
\label{vacua}
\end{eqnarray}
where $\bm\omega_0\equiv 0$ and $\bm\omega_1,...,\bm\omega_{N-1}$ are the fundamental weights of $SU(N)$.
 One can calculate the expectation value of the fundamental Polyakov  loop  at these vacua to find
\begin{eqnarray}
\label{Polyakov Vev}
\mbox{tr}_F\left[e^{i\bm \Phi_0   \cdot \bm H} \right]\bigg\vert_{\bm\Phi_0 =2\pi \bm\omega_a}=Ne^{- i\frac{2\pi a}{N}}, 
\end{eqnarray} 
where we used the fact that the trace can be expressed as a sum over the weights of the fundamental representation $\bm \nu_b$, using the formulae given in Section \ref{appendix:groups}. The nonvanishing of the Polyakov loop expectation value  (\ref{Polyakov Vev}) shows that the zero-form $\Z_N$ center symmetry is broken in the   vacua (\ref{vacua}), which are permuted by its action.\footnote{These vacua  lie  at the vertices of the affine Weyl chamber, which is defined via the inequalities $\bm \alpha_a\cdot \bm\Phi>0$ for $a=1,2,...,N-1$ and $-\bm \alpha_0\cdot \bm \Phi<2\pi$, where $\bm \alpha_0$ is the lowest, or affine, root. The $SU(N)$ gauge symmetry is unbroken at the vertices of the Weyl chamber (which can be pictured as  a triangle for $SU(3)$ and a tetrahedron for $SU(4)$), is partially broken on the faces, and completely abelianizes in the bulk of the Weyl chamber.}

We shall call a ``$k$-wall" a DW configuration in the $3$-D theory (\ref{bosonic action}), which is a (e.g.) $z$-dependent  kink  interpolating between the vacua $\bm\Phi_0 = 0$ and $\bm\Phi_0 = 2 \pi \bm\omega_k$, $k>0$. In other words, a $k$-wall satisfies the boundary conditions:
\begin{equation}
 \bm A_4^{DW}(z)=T\bm \Phi^{DW}(z),   ~~ \bm\Phi^{DW}(z \rightarrow - \infty) = 0,   ~~ \bm\Phi^{DW}(z \rightarrow + \infty) = 2 \pi \bm\omega_k, ~k>0.
 \label{kwallbc}
 \end{equation}
The $SU(N)$ gauge group is spontaneously broken on the domain wall by the nontrivial wall profile 
$\bm \Phi^{DW}(z)$,  while it gets restored at the wall boundaries $|z| \rightarrow \infty$.  A fundamental DW separates two distinct vacua, and hence, there are $C^N_2=\frac{N(N-1)}{2}$ fundamental DWs.  
 
DW ($k$-wall) configurations have been studied in the literature, both   in the high temperature limit $\beta \Lambda_{QCD} \ll 1$, where   higher loop effects have been also included \cite{Bhattacharya:1990hk, Bhattacharya:1992qb, Smilga:1993vb,KorthalsAltes:1999xb,KorthalsAltes:2000gs,Armoni:2008yp,KorthalsAltes:2009dp,Armoni:2010ny},  and on the lattice at lower temperatures \cite{Bursa:2005yv}. In particular, the  $k$-wall profiles and the $k$-wall tensions have been  studied in theories with massless adjoint fermions and scalars, such as ${\cal N}=4$ super-Yang-Mills \cite{Armoni:2008yp}, and two-index fermions \cite{Armoni:2010ny}.  

To us, the fact of crucial importance is that in the high-$T$ limit, the stable\footnote{There  exists a number of metastable DWs which can be numerically found for specific values of $N$.} $k$-wall profile takes the form
\begin{eqnarray}
 A_{4 }^{DW \; (k)}(z)  = T  Q^{(k)}(z) \tilde{H}^{N-k}\,~, 
 \label{kwallsolution}
\end{eqnarray}
where $\tilde{H}^{N-k}$ denotes the Cartan generator 
\begin{eqnarray}
\label{tildeH1}
\tilde{H}^{N-k}=\frac{1}{\sqrt{kN(N-k)}}\mbox{diag}\left[\underbrace{k,k,...,k}_{N-k\,\mbox{\small times}},\underbrace{k-N,k-N,...,k-N}_{k\,\mbox{\small times}}\right]\,,
\end{eqnarray} see also (\ref{tildeH}) in  Appendix \ref{appendix:groups}. The wall profile function  
  $Q^{(k)}(z)$ obeys the boundary conditions 
 \begin{equation}
 \label{qbc}
 Q^{(k)}(z \rightarrow - \infty) = 0,~~~~ Q^{(k)}(z \rightarrow + \infty) = - 2 \pi \sqrt{k (N -k)\over N}~.
 \end{equation} 
To obtain the solution of the $k$-wall profile, we substitute the ansatz (\ref{kwallsolution}) into (\ref{bosonic action}), taking $n_f=1$, and use the change of variables
\begin{eqnarray}
q(z)\equiv -\frac{1}{2\pi}\sqrt{\frac{N}{k(N-k)}} Q^{(k)}(z)\,, \quad z'\equiv \frac{T \sqrt{ g^2 N}}{\pi^2}z\,,
\label{change of variables}
\end{eqnarray}
along with the fact that the $(N$-$k)$-th component of the roots that contribute to the potential $V(\bm A_4)$ is given by $\gamma^{(N-k)}=\sqrt{\frac{N}{k(N-k)}}$. Then, the $k$-wall action $(n_f=1)$ reads
\begin{eqnarray}
S_{k-{\rm wall}}= 4{\cal{A}}T^2\; \frac{(N-k)k}{\sqrt{g^2 N}} \int_{-\infty}^\infty dz' \left\{\left(\frac{\partial q(z')}{\partial z'}\right)^2+\sum_{n=1}\frac{-1+(-1)^n}{n^4}\cos\left(2\pi nq(z')\right)\right\},
\label{k-wall action}
\end{eqnarray}
where ${\cal A}$ is the wall area, and  the boundary conditions (\ref{qbc}) translate into $q(z'\rightarrow -\infty)=0$ and $q(z'\rightarrow \infty)=1$. From (\ref{k-wall action}, \ref{change of variables}) it is easily seen that the $k$-wall tension follows the Casimir scaling $\frac{S_{k-{\rm wall}}}{S_{1-{\rm wall}}}=\frac{k(N-k)}{N-1}$, while its width $\sim \frac{1}{T\sqrt{g^2 N}}$ is independent of $k$.

Two comments are now in order.  First, the $k$-wall (\ref{kwallsolution}) interpolates between  the two vacua   
\begin{eqnarray} \label{kwallinfinity}
A_{4 }^{DW  \; (k) }(-\infty) &=& \mbox{diag}\left[0,0,...,0\right], \\
A_{4 }^{DW  \; (k) }(+ \infty) &=& 2 \pi T \mbox{diag}\left[\underbrace{-{k\over N},-{k\over N},...,-{k\over N}}_{N-k\,\mbox{times}}, \underbrace{1-{k\over N},1-{k\over N},...,1-{k\over N}}_{k\,\mbox{times}}\right]~.\nonumber
\end{eqnarray}
It is then easily seen that as one crosses the $k$-wall, the trace of the Polyakov loop  interpolates between $\tr e^{i \beta A_{4}^{DW \; (k)}(-\infty)} = N$  and $\tr e^{i \beta A_{4}^{DW \; (k)}(\infty)} = N e^{- i {2 \pi k \over N}}$, as in  (\ref{Polyakov Vev}), hence the $k$-wall obeys the desired boundary conditions.
 
Second, as the form of the DW profile (\ref{kwallsolution}) shows, 
 the $SU(N)$ group  breaks to $U(1)\times SU(k)\times SU(N-k)$ on the $k$-wall. The mass of the off-diagonal gauge bosons on the $k$-wall will be seen, see the following Section \ref{section:fermions} and Appendix \ref{appendix:zeromodes}, to be of order $T \sqrt{N \over k(N-k)}$. The massless gauge bosons are localized near the DW due to the fact that the bulk gauge theory has a mass gap  $\sim g^2    T$ due to $3$-D confinement in the bulk. Thus, there is an unbroken $U(1)\times SU(k)\times SU(N-k)$ $2$-D gauge theory on the $k$-wall worldvolume. 
 
As the bulk confinement scale $( g^2  T)^{-1}$ is much larger than the DW width $(T \sqrt{g^2 N})^{-1}$, the validity of the semiclassical treatment of the DW solution and the appearance of localized fermion zero modes (Section \ref{section:fermions}) is beyond doubt. 
The $k$-wall gauge coupling, however, is not precisely calculable, since the mechanism responsible for the localization of the massless gauge bosons on the wall is nonperturbative, due to bulk confinement,  as in \cite{Dvali:1996bg,Dvali:1996xe}.  As we already remarked in   \cite{Anber:2018jdf}, introducing a localization length  $\delta$ of the DW gauge fields, whose (not precisely known) value is between the DW width and the bulk confining scale,    the  strong coupling scale of the worldvolume theory is estimated as $({g^2  T\over \delta})^{1/2}$. If this scale were of the order of the bulk mass gap $g^2  T$, a 2-D QFT treatment would not be appropriate as there would be significant mixing between 3-D bulk  and 2-D DW strong coupling physics, a difficult problem awaiting a dedicated study. 
With the above remarks in mind, in what follows, we continue with a 2-D treatment of the $k$-wall theory to derive an IR 2D TQFT matching the $k$-wall 't Hooft anomalies and consistent with anomaly inflow (Sections \ref{section:mixedanomaly} and \ref{section:chiralanomaly}). We  stress, however, that our prediction of a nonvanishing bilinear condensate on the DW and the associated breaking of the $Z_{2N}^{d\chi}$ chiral symmetry (Section \ref{section:chiralanomaly}) as well as of the screening of fundamental quarks on the wall (Section \ref{Screening and strings ending on walls}) is a likely consequence of the presence of fermion zero modes, irrespective of the details of the localization of the worldvolume gauge fields. The uncertainties just discussed make a strong case  for a   lattice study, continuing  \cite{deForcrand:2004jt,Bursa:2005yv}. 
   
\subsection{Fermions and the $k$-wall worldvolume theory}
\label{section:fermions}
The $k$-wall worldvolume theory, apart form the massless $U(1) \times SU(N-k) \times SU(k)$ gauge fields, also involves the normalizable fermion zero modes in the $k$-wall background. 
Thus, we now turn to fermions on the $k$-th DW.  

We begin with introducing some necessary notation; more details are given in Appendix \ref{appendix:groups}. The unbroken $U(1)$ generator was already given in  (\ref{tildeH1}) and 
 satisfies $\mbox{tr}\left[\tilde{H}^{N-k}\tilde{H}^{N-k}\right]=1$ (we use the tilde to stress  that this is not one of the $SU(N)$ generators given in (\ref{group1})). 
Further, we break the Lie-algebra generators of $SU(N)$ as follows\footnote{For brevity, omitting $\tilde{H}^{N-k}$ of (\ref{tildeH1}), which commutes with the $SU(N-k)$ and $SU(k)$ hermitean generators  ${\cal T}^a$ and ${\cal T}^A$.}
\begin{eqnarray}\label{kwallgenerators1}
T=\left[\begin{array}{cc} {\cal T}^a_{(N-k)\times(N-k)} & E_{\bm\beta\,(N-k)\times k}\\ E_{-\bm\beta\,k\times(N-k)} & {\cal T}^A_{k\times k}  \end{array} \right]\,,
\end{eqnarray}
where the subscript indicates the matrix dimensionality. 
We expand the fermions and gauge fields using the basis of $U(1)\times SU(N-k)\times SU(k)$ generators: \begin{eqnarray}\label{kwalldecompositions}
\lambda&=& \lambda^{N-k}\tilde{H}^{N-k}+\lambda^a {\cal T}^a+\lambda^A {\cal T}^A +\sum_{CC'} \lambda^{\bm\beta_{CC'}}E_{\bm\beta_{CC'}}+\lambda^{-\bm\beta_{CC'}}E_{-\bm\beta_{C'C}}\,,\\
A_\mu&=&A_\mu^{N-k}\tilde{H}^{N-k}+A_\mu^a{\cal T}^a+A_\mu^A {\cal T}^A+\sum_{_{CC'}} A_\mu^{\bm\beta_{CC'}} E_{\bm\beta_{CC'}}+A_\mu^{-\bm\beta_{CC'}} E_{-\bm\beta_{CC'}}\,,
\end{eqnarray} %
where the sums over $C$ and $C'$ run over $1,...,N-k$ and $1,...,k$, respectively;  for brevity, the ranges of these sum as well as those over $a$ (the $SU(N-k)$ generators) and $A$ (the $SU(k)$ generators) are not explicitly shown.

We now note that $A_\mu^{N-k}$ includes the $k$-wall background (\ref{kwallsolution}). The first commutation relation in (\ref{HE}) then implies that the ``$W$-bosons" $A_\mu^{-\bm \beta_{CC'}}$ (the gauge field component  along the broken generators $E_{\pm \bm\beta_{CC'}}$) obtain mass of order $T \gamma^{(N-k)} = T \sqrt{N \over k(N-k)}$ on the $k$-wall, as already noted. Thus, we  ignore the $W$-boson fields in what follows.
Further,  the behavior of the fermions is determined by their covariant derivative $D_\mu \lambda =\partial_\mu \lambda -i[A_\mu,\lambda]$. From (\ref{kwalldecompositions}) and the fact that the DW background commutes with $\tilde{H}^{N-k}$, ${\cal T}^a$, and ${\cal T}^A$, it follows that the fields $\lambda^{N-k}$, $\lambda^a$, and $\lambda^A$ do not couple to the DW. These fields  would remain massless, were it not for the antiperiodic boundary conditions associated with the compact Euclidean time direction, which give them a $3$-D mass of order $T$. Since they do not couple to the DW, they remain massive in the $k$-wall background and we also ignore them in what follows. 
 \begin{table}[h]
\begin{center}
\begin{tabular}{|c|c|c|c|c|c}
\hline
fermion field &  $\psi_+$ & $\psi_-$\\\hline 
$2$-D chirality  & left mover & right mover\\\hline \hline
gauge $U(1)$ & $\gamma^{(N-k)} \equiv \sqrt{\frac{N}{k(N-k)}}$ & $-\gamma^{(N-k)}$\\\hline
gauge $SU(k)$ & $\overline\Box$ & $\Box$\\\hline
 gauge $SU(N-k)$ & $\Box$ & $\overline\Box$\\  \hline \hline
global $U(1)_{R}$ & 1 & 1 \\ \hline
\end{tabular}
\caption{ \label{charges of k DW1}The massless fermions  of the $k$-wall worldvolume theory and their charges under the $U(1)\times SU(N-k) \times SU(k)$ gauge group and the bulk global $U(1)_R$ chiral symmetry. Opposite chirality fermions are in conjugate representations, thus the $2$-D $k$-wall theory is axial, while the bulk $U(1)_R$ chiral symmetry is  vectorlike on the $2$-D worldolume. A $\Z_{2N}$ subgroup of $U(1)_R$ is anomaly free on the $k$-wall worldvolume, as in the $4$-D bulk, see (\ref{jacobian1}, \ref{discretechiral}).}
\end{center}
\end{table}

Since,  as explained above,  all other fermions are massive,  the object of our  interest is the coupling of the  zero modes of the $\lambda^{\bm \beta_{CC'}}$ fermions    to the massless gauge fields on the wall. The detailed derivation is given in Appendix \ref{appendix:zeromodes}. Here we just summarize 
the resulting $k$-wall worldvolume theory: it has massless 
 $U(1)\times SU(N-k)\times SU(k)$ gauge fields and fermions $\psi_+$ and $\psi_-$ with quantum numbers given in Table \ref{charges of k DW1}. The matter part of the Lagrangian of the $k$-wall theory can be written as:
 \begin{eqnarray}\label{matterkwalllagrangian}
 L_{k-{\rm wall}} &=& i \;\tr \bar\psi_+ (  \partial_- \psi_+ - i \gamma^{(N-k)} A_-^{N-k} \;\psi_+ - i A_-^a {\cal T}^a \;\psi_+  + i \psi_+ \;A_-^A {\cal T}^A ) \nonumber \\ 
 &+&   i\; \tr \bar\psi_- (\partial_+ \psi_-+ i \gamma^{(N-k)} A_+^{N-k} \;\psi_-- i A_+^A {\cal T}^A\;\psi_-  + i \psi_- \;A_+^a {\cal T}^a ) ~,
 \end{eqnarray}
 where $\psi_+$ is represented as a $(N-k) \times k$ matrix and $\psi_-$ as a $k \times (N-k)$ matrix.
The $SU(N-k)$ and $SU(k)$ generators ${\cal T}^a$, ${\cal T}^A$ are the ones from (\ref{kwallgenerators1}) and $\partial_\pm = \partial_1 \pm i \partial_2$.
 
 In addition to the zero-form symmetries discussed above and shown in Table \ref{charges of k DW1}, the $k$-wall theory inherits the reduction of the  $\Z_{N}^{(1)}$ 1-form global symmetry of the underlying $SU(N)$ bulk theory to the $2$-D worldvolume. Its action on the transition functions for the gauge fields on the torus is given in Appendix \ref{appendix:quantization}.
 
 \subsection{Anomalies on the $k$-wall and anomaly inflow}
 
The two-dimensional anomaly-free axial theory (\ref{matterkwalllagrangian}) has a classical global (vectorlike) $U(1)_R$ symmetry, where $\psi_\pm$ have the same charge, as per  Table \ref{charges of k DW1}. This symmetry is inherited from the classical bulk chiral $U(1)_R$ symmetry. Recall that in the $4$-D bulk $SU(N)$ theory the chiral anomaly breaks $U(1)_R \rightarrow \Z_{2 N}^{d \chi}$. Similarly, the $2$-D vectorlike global $U(1)_R$ of Table \ref{charges of k DW1} is anomalous. There is no $2$-D mixed $U(1)_R$-$SU(N-k)$ or $U(1)_R$-$SU(k)$ anomaly, but only a $U(1)_R$-$U(1)$ anomaly. 
Under a $U(1)_R$ transformation, $\psi_\pm \rightarrow e^{i \chi} \psi_\pm$, the $2$-D fermion measure, denoted by ${\cal D} \psi$, changes as\footnote{The factor of $2$ in the exponent occurs because the $2$-D left- and right- movers $\psi_+$ and $\psi_-$ have opposite signs of the Jacobian, but also opposite gauge-$U(1)$  charges, while  the $(N-k)k$  factor counts the number of charged fermion components.}
\begin{equation}
\label{anomaly$2$-D0}
{\cal D} \psi \rightarrow {\cal J}\; {\cal D} \psi , ~{\rm where} ~~ {\cal J} \equiv \exp\left[ i \; 2 \chi (N-k)k \; \gamma^{(N-k)} \oint {F_{12}^{N-k} dx^1 dx^2   \over 2\pi} \right],
\end{equation}
where $\oint {F_{12}^{N-k} dx^1 dx^2 \over 2\pi}$ is the $U(1)$ flux through the $2$-D torus (as usual, to study anomalies, we imagine that the $k$-wall plane is compactified to a two-torus  $x^1 \in (0,L_1]$ and $x^2 \in (0,L_2]$). 

In order to determine the anomaly-free chiral symmetry, we need to understand the $U(1)$ flux quantization. This entails understanding   the  boundary conditions for the $U(1) \times SU(N-k) \times SU(k) \in SU(N)$ gauge bundle on the torus, a question addressed in Appendix \ref{appendix:quantization}. There,  we show that 
in the $SU(N)$ theory  the $U(1)$ flux is quantized in units of $\gamma^{(N-k)}$ 
\begin{eqnarray}
\label{solution2}
\oint {F_{12}^{N-k} dx^1 dx^2 \over 2\pi} \;   &=&   \gamma^{(N-k)} \;  n ,~ n \in \Z.
\end{eqnarray}
 A physical way to interpret this quantization condition is as follows.  A fundamental of $SU(N)$ decomposes  into two representations  under the unbroken $U(1) \times SU(N-k) \times SU(k)$ gauge group: $q_1 \sim ({k \over N} \gamma^{(N-k)}, \Box, {\bf 1})$ and $q_2\sim (({k -N \over N}) \gamma^{(N-k)}, {\bf 1}, \Box)$, as seen from (\ref{tildeH1}, \ref{kwallgenerators1}).  The $SU(N-k)$--singlet ``baryons" $(q_1)^{N-k}$ and their $SU(k)$ counterparts $(q_2)^k$  both have the same absolute value of $U(1)$ charge $1/\gamma^{(N-k)}$. The flux quantization condition (\ref{fluxes7}) is precisely the one  appropriate for particles of charge $1/\gamma^{(N-k)}$. The condition (\ref{solution2}) is also discussed in Section \ref{section:constantflux} using constant flux backgrounds and derived from considering the  boundary conditions on the $2$-D torus in  Appendix \ref{appendix:quantization}.

 Substituting (\ref{solution2}) into the measure transformation (\ref{anomaly$2$-D0}) we find that the Jacobian of a $U(1)_R$ transformation is
\begin{eqnarray}
\label{jacobian1}
{\cal J} = e^{ 2 i \chi N n}~.
\end{eqnarray}
 The anomaly-free subgroup of $U(1)_R$ is determined by the condition that ${\cal J}=1$ for all $n$, hence $\chi = {2\pi \over 2N}$ gives a unit Jacobian and there is an anomaly free $\Z_{2 N}^{d \chi} \in U(1)_R$ discrete symmetry on the $k$-wall worldvolume---inherited from the bulk anomaly free chiral symmetry. As the $2$-D $k$-wall theory is axial, the anomaly free subgroup of   $U(1)_R$ is vectorlike:
 \begin{equation}\label{discretechiral}
 \Z_{2N}^{d \chi}: \psi_\pm \rightarrow e^{ i {\pi \over N}} \psi_\pm~.
 \end{equation}
 
Before we continue the discussion of anomalies, we pause and, in the following Section \ref{section:constantflux}  give a perhaps more transparent derivation of (\ref{solution2}), making use of a particular constant flux background; a more formal derivation is in Appendix \ref{appendix:quantization}. The reader interested in the mixed zero-form/one-form anomaly can proceed to Section \ref{section:mixedanomaly}.
 
 \subsubsection{Flux quantization from a constant flux background}
\label{section:constantflux}
Here, we consider constant field strength backgrounds on the $2$-D torus, which can  be rotated into the Cartan subalgebra. 
The constant field strength background we use  here to motivate the flux quantization (\ref{solution2}) (and (\ref{fluxes4}) below) is an example of configurations obeying the twisted boundary conditions discussed in the Appendix, see (\ref{sunembeddedccocycle}).\footnote{For example, the background (\ref{fluxes1}) below with $\bm u = \bm\alpha^{N-k}$  is a gauge field configuration of the $SU(N)$ theory that is  summed over in the $k$-wall theory path integral.}
We denote the gauge field in these flux backgrounds ${\cal A}_i$  (here and below $i=1,2$, $\epsilon_{12}=1$) and take
\begin{eqnarray}
\label{fluxes1}
{\cal A}_i(x_1,x_2)&=& {\pi \; n_{12}\; \epsilon_{ij}\; x_j \over L_1 L_2}\; {\bm u} \cdot {\bm H}~, 
\end{eqnarray}
where $n_{12}$ is an integer, ${\bm u}$ is a vector in the Cartan subalgebra of $SU(N)$ whose possible values will be discussed shortly, and $\bm H$ are the $SU(N)$ Cartan generators defined in (\ref{group1}). For unconstrained $\bm u$, (\ref{fluxes1}) represents general constant field strength (${\cal F}_{12} = \partial_1 {\cal A}_2 - \partial_2 {\cal A}_1$) backgrounds.  
 The  gauge backgrounds (\ref{fluxes1}) are   periodic  up to gauge transformations $\Omega_{1,2}$: 
\begin{eqnarray}
\label{fluxes2}
{\cal A}_i(x_1+L_1, x_2)&=&\Omega_1(x_2)[{\cal A}_i(x_1,x_2) + i   \partial_i ]\Omega_1^\dagger(x_2),    ~~\Omega_1(x_2)\equiv e^{  i {\pi n_{12} x_2 \over L_2} \;{\bm u} \cdot {\bm H} }, \nonumber\\
{\cal A}_i(x_1, x_2+L_2)&=&\Omega_2(x_1)[{\cal A}_i(x_1,x_2) + i  \partial_i ]\Omega_2^\dagger(x_1),  ~~\Omega_2(x_1)\equiv e^{-i {\pi n_{12} x_1 \over L_1}\; {\bm u} \cdot {\bm H}}.
\end{eqnarray}
The matrices $\Omega_{1,2}$  are the transition functions of a $SU(N-k)\times SU(k) \times U(1)$  bundle on the torus \cite{tHooft:1979rtg,tHooft:1981sps,vanBaal:1982ag} and  obey a consistency condition at the corners of the torus, the cocycle condition, which reads ($z$ in (\ref{fluxes3}) is a $\Z_N$ phase): 
\begin{eqnarray}
\label{fluxes3}
\Omega_1(L_2) \Omega_2(0) &=& z\;\Omega_2(L_1) \Omega_1(0) \\
\implies  && e^{i 2 \pi n_{12}  {\bm u} \cdot {\bm H}} = z~\implies \left\{ \begin{array}{cc} \bm u \in \{\bm \alpha_1, ..., \bm \alpha_{N-1}\}, &{\rm if}\; z =1, \\\bm u \in \{\bm \omega_1, ..., \bm \omega_{N-1}\}, &{\rm if}\; z \in \Z_N,   z\ne 1, \end{array}  \right. \label{fluxes4}  
\end{eqnarray}
where we used the specific form of $\Omega_{1,2}$ from (\ref{fluxes2}) in (\ref{fluxes4}).  Notice that   only the product of the $SU(N-k)$, $SU(k)$, and $U(1)$ transition functions  $\Omega_{1,2}$ corresponding to  the constant flux background (\ref{fluxes1}), but not the individual ones, obeys the cocycle condition (\ref{fluxes3}) (recall the earlier remark from footnote \ref{groupfootnote} that the gauge group of the $k$-wall theory is  $SU(N-k)\times SU(k) \times U(1)/\left(\Z_{N-k} \times \Z_k\right)$). In the $SU(N)$ theory, only $z \equiv 1$ is allowed, hence $\bm u$ should be an element of the $SU(N)$ root lattice, recall (\ref{root weights}), i.e. $\bm u = \bm \alpha_a$, any of the simple roots, or an  integer valued linear combination thereof.\footnote{To see this recall from Appendix \ref{appendix:groups} that $\bm\alpha_a = \bm\nu_a- \bm\nu_{a+1}$ and that $\bm\nu_1,...,\bm \nu_N$,  are the eigenvalues of $\bm H$ (\ref{group1}). 
 Thus $e^{i 2\pi \bm\alpha_a \cdot \bm H}$ is a diagonal matrix with eigenvalues $e^{i 2 \pi (\bm \nu_a - \bm \nu_{a+1}) \cdot \bm \nu_c } = 1$ for all $a, c$. } 
 On the other hand, nontrivial $z$ factors describe 't Hooft fluxes in the $SU(N)$ theory, i.e.~nontrivial two-form center symmetry gauge backgrounds. Thus, a generic 't Hooft flux background also permits $\bm u =\bm\omega_k$, for any weight vector $\bm\omega_k$, so that  $\bm u$ is an element in the weight lattice as indicated in (\ref{fluxes4}). The backgrounds with $\bm u = \bm \omega_k$ are considered explicitly in Appendix \ref{appendix:thooftprojection}. In this section, we are after the $U(1)$-flux quantization in the $SU(N)$ theory and consider in detail the $\bm u = \bm \alpha_k$ case.
 
We now compute the field strength flux of  the background (\ref{fluxes1}) through the torus  
\begin{eqnarray}
\label{fluxes5}
 \oint {{\cal F}_{12} dx^1 dx^2 \over 2 \pi}   &=& - n_{12} \; {\bm u} \cdot {\bm H}~. 
\end{eqnarray}
It is clear from (\ref{fluxes4}) that the eigenvalues of (\ref{fluxes5}) are integers for $\bm u = \bm\alpha_a$ (i.e. $\bm u$ in the root lattice) and are valued in $\Z\over N$ for $\bm u = \bm \omega_a$ (i.e. $\bm u$ in the weight lattice).
We now project the $SU(N)$ Cartan subalgebra flux  (\ref{fluxes5}) onto the $\tilde{H}^{N-k}$ generator (\ref{tildeH1}) (as this is the only part of the $SU(N)$ flux appearing in the Jacobian of the $U(1)_R$ transformation (\ref{anomaly$2$-D0})). 
To this end we note that the generators $H^a$ of (\ref{group1}) form a complete orthonormal set of traceless diagonal $N\times N$ matrices; a different orthonormal set can be found, which   includes the unit-norm (\ref{tildeH}) generator $\tilde{H}^{N-k}$ as one of its elements. Thus,  the projection of $H^a$ (\ref{group1}) on $\tilde{H}^{N-k}$ (\ref{tildeH}) is 
\begin{eqnarray}\label{projection1}
\tr   H^a \tilde{H}^{N-k} &=& -{\sqrt{N\over k  (N-k)}}  \sum\limits_{A=N-k+1}^N \lambda^{aA}   ~.
\end{eqnarray}

Our interest is really in the projection of $\bm u \cdot \bm H$ onto $\tilde{H}^{N-k}$. Thus, we find that (\ref{fluxes5}), projected on $\tilde{H}^{N-k}$ equals (recalling from (\ref{weightsoffund}) that $\lambda^{aA} = (\bm \nu^A)^a$)

\begin{eqnarray} \label{fluxes6}
\oint { {F}_{12}^{N-k} dx^1 dx^2 \over 2 \pi}  &=& - n_{12} \; {\bm u} \cdot \tr {\bm H} \tilde{H}^{N-k}~ \\
&=&
n_{12} \sqrt{N \over k(N-k)} \sum\limits_{A=N-k+1}^N \sum\limits_{a=1}^{N-1} (\bm \nu^A)^a  ({\bm u})^a =   n_{12}\; \gamma^{(N-k)} \sum\limits_{A=N-k+1}^N \bm \nu^A \cdot \bm u~. \nonumber
\end{eqnarray}
We recall from (\ref{root weights}, \ref{betaroots}) that the roots are differences of fundamental weights, thus   $ \bm \nu^A \cdot \bm \alpha^a = \bm \nu^A \cdot \bm \nu^a - \bm \nu^A \cdot \bm\nu^{a+1} = \delta^{A,a} - \delta^{A,a+1}$; thus, the sum $\sum_{A=N-k+1}^N   \bm \nu^A \cdot \bm \alpha^a$ appearing in (\ref{fluxes6}) is $0$ unless $a = N-k$ when it equals $-1$.
Finally, we obtain the $U(1)$-flux quantization condition (\ref{fluxes5}, \ref{fluxes6}) in the form
\begin{equation}
\label{fluxes7}
\oint { {F}_{12}^{N-k} dx^1 dx^2 \over 2 \pi}  = \gamma^{(N-k)} \times n~, ~ n\in \Z,
\end{equation}
which agrees with the one quoted earlier in (\ref{solution2}); see also the more general discussion in Appendix \ref{appendix:quantization}.

\subsubsection{Mixed discrete chiral-center anomaly}
\label{section:mixedanomaly}
 
Backgrounds for the  discrete $\Z_N^{(1)}$ one-form center symmetry are nontrivial 't Hooft fluxes of the $SU(N)$ theory. The corresponding boundary conditions in the $U(1) \times SU(N-k) \times SU(k)$ theory are studied in Appendix \ref{appendix:quantization}, where the general rule for $U(1)$ flux quantization in a nontrivial $SU(N)$ 't Hooft flux is found. Explicit constant flux examples are given in Appendix  \ref{appendix:thooftprojection}.  Introducing nontrivial topological backgrounds for the one-form symmetry is equivalent to introducing  't Hooft fluxes, labeled by $p \in \Z$ (mod $N$).

Consider now the fate of an anomaly-free $\Z_{2N}$ chiral symmetry transformation (\ref{discretechiral}). The measure transforms with a Jacobian (\ref{anomaly$2$-D0})
\begin{equation}
\label{jacobianthooft}
{\cal J} = e^{i {2 \pi \over N} k (N-k) \gamma^{(N-k)} \alpha }\,,
\end{equation}
where $\alpha$ denotes the $U(1)$ flux, $\alpha= \oint {F_{12}^{N-k} dx^1 dx^2 \over 2\pi}$, see eqn.~(\ref{u1fluxsuN}). The solution for $\alpha$ for a general nonzero  't Hooft flux\footnote{In Appendix \ref{appendix:quantization}, the $\Z_N$ twist corresponding to nontrivial 't Hooft flux is denoted by $p_N =1,...,N-1$.} $p \equiv p_N$,  from (\ref{solution34}), is given by $\alpha \gamma^{(N-k)} k (N-k) =   p (N-k) - N p_{N-k} - N(N-k) m_4$. 
Substituting into (\ref{jacobianthooft}), we obtain a nontrivial Jacobian of the $\Z_{2N}^{d \chi}$ transformation in the 't Hooft flux background
\begin{equation}
\label{jacobianthooft1}
{\cal J}= e^{i {2 \pi \over N} p (N-k)} = e^{- i {2\pi \over N} k p}~.
\end{equation}
We conclude that the $k$-wall theory has a 't Hooft anomaly between the $\Z_{2N}$ discrete chiral symmetry and the 1-form $\Z_N$ center symmetry of the $4$-D theory projected on the DW plane.
  
Also, for further use (Section \ref{section:chiralanomaly}),  note that the effect of turning on a single unit  of 't Hooft flux in the $x^1$-$x^2$ plane has the effect of   turning on $k$ units of fractional (recall the $U(1)$ quantization condition (\ref{solution2})) $U(1)$ flux $-{\gamma^{(N-k)} \over N}$ in the $k$-wall worldvolume theory: eq.~(\ref{jacobianthooft1}) follows from (\ref{jacobianthooft}) with $\alpha =   k {p} ({-{\gamma^{(N-k)} \over N}})$. One way to physically understand this is that the $k$-wall can be thought of as the result of the merging of $k$ $1$-walls into the minimal action configuration, with each of the $k$ $1$-walls contributing equally to the total anomaly, thus multiplying the result by $k$.  

The appearance of the extra factor of $k$ in the phase of the Jacobian for the $k$-wall is also naturally expected from the anomaly inflow argument. The $\Z_{2N}^{d \chi}$-($\Z_N^{(1)}$)$^2$ anomaly in the $4$-D theory is  the variation of a 5d Chern-Simons term:
 \begin{equation}
 \label{5dCS1}
 S_{5-D} =   i \;{2 \pi \over N} \int\limits_{M_5 \;(\partial M_5 = M_4)}  {2 N A^{(1)} \over 2 \pi} \wedge
  {N B^{(2)} \over 2 \pi} \wedge  {N B^{(2)} \over 2 \pi} ~, 
  \end{equation}
  such that the $4$-D spacetime $M_4$ is the boundary of $M_5$. Here $A^{(1)}$ and $B^{(2)}$ are 1-form and 2-form gauge fields, respectively,  gauging the $\Z_{2N}^{d \chi}$ 0-form chiral  and $\Z_N^{(1)}$   center symmetries of the $4$-D theory. As in  \cite{Kapustin:2014gua}, they are defined as pairs: for the discrete chiral   $\Z_{2N}^{d \chi}$, we have $(A^{(1)}, A^{(0)})$: $2 N A^{(1)} = d A^{(0)}$ ($\oint A^{(0)} \in 2 \pi \Z$, so that
 $e^{i \oint A^{(1)}} = e^{ i {2 \pi \over 2 N} \Z}$), while for the $\Z_N^{(1)}$   center symmetry $(B^{(2)}, B^{(1)})$ obey $N B^{(2)} = d B^{(1)}$ ($\oint  B^{(1)} \in 2 \pi \Z$, so that
 $e^{i \oint B^{(2)}} = e^{ i {2 \pi \over N} \Z}$), where the integrals are over closed 1- and 2-cycles as appropriate. Under chiral symmetry $\delta_{\Z_{2N}} A^{(1)} = d \phi^{(0)}$, $\oint d \phi^{(0)} \in 2 \pi Z$, so  the closed $A^{(1)}$ Wilson loop is invariant.\footnote{For use below, under center symmetry we have $B^{(1)} \rightarrow N   \lambda^{(1)}$, $B^{(2)}\rightarrow d \lambda^{(1)}$  with $\oint d \lambda^{(1)}  \in 2 \pi \Z$, so that $e^{i \oint B^{(2)}}$ is gauge invariant (and, as already mentioned, valued in  $e^{ i {2 \pi \over N} \Z}$)  
 \cite{Kapustin:2014gua}.\label{oneform}} 
 
 Then, 
under a $\Z_{2N}$ chiral symmetry transformation with parameter $\phi^{(0)}\vert_{M_4} = {2 \pi \over 2N}$,  the variation of the Chern-Simons action (\ref{5dCS1}) localizes to the physical boundary $M_4$  \begin{equation}
 \label{5dCS2}
 \delta_{\chi} S_{5-D} =   i\; {2 \pi \over N}\; {2 N \phi^{(0)}\vert_{M_4} \over 2 \pi} \; \int\limits_{M_4  } 
  {N B^{(2)} \over 2 \pi} \wedge  {N B^{(2)} \over 2 \pi} ~=   i {2 \pi \over N} \; m,
  \end{equation}
and is equal to the   variation of the phase of the $4$-D partition function under a discrete chiral symmetry in a nontrivial  't Hooft flux background, where $\int\limits_{M_4  } 
  {N B^{(2)} \over 2 \pi} \wedge  {N B^{(2)} \over 2 \pi} = m \in \Z$ is nonzero.
  
Turning on a $B^{(2)}$ background $\oint_{M_{x^3 x^4}} {B^{(2)} N \over 2 \pi} = k$  corresponds to $k$ units of  't Hooft flux in the ${x^3}$-$x^4$ plane denoted by $M_{x^3 x^4}$ ($x^4$ is the compact time direction). In the center broken high-$T$ phase, this induces a $k$-wall configuration with worldvolume perpendicular to $x^3$ and separating two center-breaking vacua.\footnote{This procedure is equivalent to imposing twisted boundary conditions and  has been  used in lattice simulations  \cite{Bursa:2005yv}. The $k$-wall is the minimum action configuration in the background with $k$ units of 't Hooft flux. A stack of $k$ $1$-walls   also obeys the boundary conditions but has higher action (recall the Casimir scaling (\ref{k-wall action})).} In this background, the 5d CS term reduces to a $3$-D one, with $\partial M_3 = M_2$, the $k$-wall world volume: 
 \begin{equation}
 \label{3-DCS1}
 S_{3-D} =   i \;{2 \pi k \over N} \int\limits_{M_3 \;(\partial M_3 = M_2)} { 2 N A^{(1)} \over 2 \pi} \wedge
  {N B^{(2)} \over 2 \pi}   ~.
  \end{equation}
The $\Z_{2N}^{d \chi}$   variation of $S_{3-D}$ localizes to the $k$-wall worldvolume and is given by
\begin{equation}
\label{3-DCS21}
  \delta_{\chi} S_{3-D} =   i \;{2 \pi k \over N} \; {2 N \phi^{(0)}\vert_{M_2} \over 2 \pi} \; \int\limits_{M_2  } 
  {N B^{(2)} \over 2 \pi}  ~=   i {2 \pi k p\over N},
  \end{equation}
where, in the last equality, we turned on $p$ units of 't Hooft flux in the $12$ plane of the $k$-wall $\int_{M_2  } 
  {N B^{(2)} \over 2 \pi} = p$, as in obtaining (\ref{jacobianthooft1}). 
The variation (\ref{3-DCS21}) of the 3-D Chern-Simons ``anomaly inflow" term (\ref{3-DCS1}) is equal to the one obtained from the $k$-wall theory.
 
 \section{Screening and strings ending on walls}
 \label{Screening and strings ending on walls}
 
To probe the  confinement properties of the $k$-wall theory (\ref{matterkwalllagrangian}), we turn to the behavior of Wilson loops.   As already noted, a fundamental of $SU(N)$ decomposes  into two representations  under the unbroken $ U(1)\times SU(N-k) \times SU(k)$ gauge group: 
\begin{equation} q_1 \sim ( {k \over N} \gamma^{(N-k)}, \Box, {\bf 1}), ~~ q_2\sim (  ({k -N \over N}) \gamma^{(N-k)}, {\bf 1}, \Box). \end{equation}  
Further, the trace of an $SU(N)$-fundamental Wilson loop,  $W_{SU(N)}$, when reduced to the massless sector of the $k$-wall theory,\footnote{When considering the worldvolume theory in isolation, one could also  introduce separate Wilson loops for the three $k$-wall gauge groups; however, these loops do not probe the center symmetry of the bulk $SU(N)$ theory.}  becomes 
\begin{eqnarray}
\label{wilsonloop12} 
W_{SU(N)} \simeq W_{q_1} + W_{q_2}~.\end{eqnarray}
Explicit expressions for the Wilson loops $W_{q_1}$ and $W_{q_2}$, for definiteness taken to wind  in the $x^1$ direction of the $k$-wall wordlvolume, are
\begin{eqnarray}\label{wilsonloop2}
W_{q_1} (x_2) &=& \tr\left[ e^{i \int\limits_{ 0}^{L_1} A_{1 \; [N-k]}   dx^1} \Omega_{1 \; [N-k]} (x_2)\right]\; \; \; e^{i {k \over N} \gamma^{(N-k)}  \int\limits_{ 0}^{L_1} A_1^{N-k}   dx^1}e^{i \omega_{1} (x_2) 
 {k \over N} \gamma^{(N-k)}}, \nonumber \\
 W_{q_2}(x_2) &=& \tr\left[ e^{i \int\limits_{ 0}^{L_1} A_{1 \; [k]}   dx^1} \Omega_{1 \; [k]} (x_2)\right] \;\; \;  e^{i {k -N\over N} \gamma^{(N-k)}  \int\limits_{ 0}^{L_1} A_1^{N-k}   dx^1}e^{i \omega_{1} (x_2) 
 {k -N\over N} \gamma^{(N-k)}} ~,
 \end{eqnarray}
where, as described in Appendix \ref{appendix:quantization},  to insure gauge invariance we  inserted appropriate $U(1), SU(N-k)$, and $SU(k)$ transition functions $e^{i \omega_1}$, $ \Omega_{1 \; [N-k]}$, $\Omega_{1 \; [k]}$ . 
Under a $\Z_{N}^{(1)}$ center symmetry transformation, see (\ref{znoneformaction}, \ref{zn}),  in the  $i=1$ direction, the Wilson loops (\ref{wilsonloop2}) transform as $W_{q_1\; (q_2)} (x_2)$ $\rightarrow e^{i {2 \pi \over N} p_{(1)  }}   W_{q_1 \; (q_2)} (x_2)$, as appropriate for a $\Z_N^{(1)}$ 1-form symmetry. 

It has been argued a long time ago \cite{Gross:1995bp} that nonabelian  gauge theories with massless fermions in $2$-D are in the screening rather then the confining phase. One argument for screening in a massless adjoint theory is based on the equivalence of the effective actions (or fermion determinants, which are exactly calculable in $2$-D) for massless  Majorana adjoint fermions to that of  $N$-fundamental Dirac flavors. Since the latter screen fundamental charges, the equivalence of the effective actions implies that the adjoint theory also screens, i.e. breaks its 1-form center symmetry. For more general theories with  massless  fermions, one can use the observation of \cite{Armoni:1997ki,Armoni:1998kv,Armoni:1999xw} that the effect of an external source in any representation of the gauge group can be removed by a judiciously chosen chiral rotation of the fermions. This argument also holds for our  $U(1) \times SU(N-k) \times SU(k)$  $k$-wall theory  with massless left-moving fermions $\psi_+ \sim (\gamma^{(N-k)}, \Box, \overline\Box)$ and right-moving fermions in the conjugate representation. The screening also holds for the simplest case of $k=1$ walls in an $SU(2)$ gauge theory, where the worldvolume theory is abelian, see \cite{Anber:2018jdf}. 

  The fact that   fundamental charges are screened on the $k$-wall means that confining strings can end on these hot DW. Consider an $N$-ality $p$ flux due to a probe quark in the bulk. As the flux approaches the $k$-wall, due to the Higgsing of the gauge group on the wall, the flux reduces  to a $U(1) \times SU(N-k) \times SU(k)$ flux, which is  screened by the wall's massless fermions, allowing thus the flux tube to end on the DW.
This effect is interesting from several points of view. 

To the best of our knowledge,    it was first  observed in the strong coupling limit of  $T>0$ ${\cal N}=4$ super-YM (we consider the infinite spatial volume limit) via holography \cite{Aharony:1998qu}. There, the  deconfined phase DW  are represented by  Euclidean $D1$-branes on which fundamental strings  can end (see  \cite{Armoni:2008yp,Armoni:2010ny} for discussions of $k$-walls with $k\sim N/2$). In this paper, we found a semiclassical explanation of this  in supersymmetric (as well as nonsupersymmetric) YM theory with massless adjoints, based on the screening properties of the $2$-D DW theories containing massless fermions. We note that our semiclassical findings also apply to the case of  $T>0$ deconfined phase of weakly coupled ${\cal N}=4$ super-YM.\footnote{The only difference is the number of massless fermions. There is also no discrete chiral symmetry in ${\cal N}=4$ SYM (as it is broken by various Yukawa couplings) but only a nonabelian $SU(4)$  flavor ($R$-) symmetry, as in QCD(adj) with $n_f=4$.} 
 
 The second  observation is about the intriguing similarities between the $2$-D physics on the high-$T$ DW and  the physics on the $3$-D (or, sometimes, $2$-D, see below)  DW associated with the broken discrete chiral $R$-symmetry  in the low-$T$ confined phase of super-YM theory.  That confining strings can end on these low-$T$ domain walls was shown first using the $M$-theory embedding (in the essentially $\R^4$ setup of    \cite{Witten:1997ep}, the worldvolume of these DW is $3$-D).  Recently, such behavior has also been explained semiclassically, using only weakly coupled semiclassical quantum field theory arguments, in the low temperature phase of super-YM and QCD(adj) on $\R^3 \times \S^1$ \cite{Anber:2015kea}. Here, the DW worldvolume is $2$-D, similar to the high-$T$ domain walls discussed in this article.  In both cases,  quarks are deconfined and therefore the one-form bulk center symmetry is broken on the worldvolume of these DW. In the calculable $\R^3 \times \S^1$  setup, the physics of deconfinement on the walls is quite explicit and well understood, especially in the case of $k=1$ walls between neighboring chiral-broken vacua (the semiclassical understanding of $k>1$ DW between $R$-symmetry breaking vacua on $\R^3 \times \S^1$ is not yet complete); see Figure \ref{fig:01} for illustration.
Achieving a microscopic quantum field theory understanding of the  mechanism leading to  deconfinement on the $3$-D walls in   $\R^4$    \cite{Acharya:2001dz} and of its relation to that on $k=1$ walls on $\R^3\times\S^1$, as understood in  \cite{Anber:2015kea},  and  on general $k>1$ walls,   would be of interest.\footnote{We thank Zohar Komargodski for discussions of unpublished work on this topic.}
 
\section{Discrete chiral symmetry and the IR matching of the anomaly}
\label{section:chiralanomaly}

In order to answer the question of how the mixed discrete-chiral/center anomaly of Section \ref{section:mixedanomaly} is matched by the IR physics of the $k$-wall theory, we need to be cognizant of the IR behavior of the $2$-D worldvolume theory given in Table \ref{charges of k DW1} and 
eq.~(\ref{matterkwalllagrangian}). As opposed to the $SU(2)$ case  \cite{Anber:2018jdf}, where the worldvolume theory was exactly solvable, we can not rigorously show how the theory behaves. However, arguments involving nonabelian  bosonization and gauged Wess-Zumino-Witten (WZW) models, along the lines of \cite{Frishman:1992mr,Frishman:2010zz},   suggest that the $k$-wall $U(1) \times SU(N-k) \times SU(k)$ theory  develops a nonvanishing bi-fermion condensate (in SYM). 

The nonabelian bosonization \cite{Witten:1983ar} is a set of rules that map fermionic to bosonic operators, see \cite{Frishman:1992mr,Frishman:2010zz} for reviews. Using these rules one can show that the action of $N$ free Majorana fermions, which is invariant under some global symmetry $G$, is equivalent to a WZW model of a nonabelian bosonic field ${\cal U}$, which is a matrix in $G$. One can also gauge an appropriate $H \subset G$, which yields $2$-D QCD with gauge group $H$ and fermions in the fundamental representation of $H$. The gauge theory is then  mapped to a gauged version of WZW model. To be more specific, we consider  the worldvolume theory of the $k$$=$$1$-wall, which, from (\ref{matterkwalllagrangian}),  is   $2$-D QCD with gauge group\footnote{To avoid confusion, note that the correlators in the following paragraph  refer to the vectorlike version of the axial worldvolume theory, obtained by relabeling $\psi_+ \leftrightarrow \bar\psi_+$ in  (\ref{matterkwalllagrangian}).}  $U(1)\times SU(N-1)$. It was argued that the bosonization rule for the fermion bilinear $\bar\psi_+ \psi_-$ in this theory is given by
\begin{eqnarray}
\bar\psi_+^{a}\psi_{-b}=\mu\, h^{a}_b \, e^{-i\sqrt{\frac{4\pi}{N-1}}\phi}\,,
\label{bosonization rule}
\end{eqnarray}
where $\mu$ is a normalization scale and $h$ and $e^{-i\sqrt{\frac{4\pi}{N-1}}\phi}$ are   bosonic fields, $SU(N-1)$ and $U(1)$ group elements, respectively. In the gauged $U(1)\times SU(N-1)$ theory, if the fermions are very light or massless (as is the case in our worldvolume theory),  the   $h$ and $\phi$ sectors of the theory become strongly coupled and acquire  a mass gap. The correlators $ \langle e^{-i\sqrt{\frac{4\pi}{N-1}}\phi(x)}e^{ i\sqrt{\frac{4\pi}{N-1}}\phi(y)}\rangle $ and $\left\langle \mbox{tr}  h^\dagger(x) \;\mbox{tr} h(y) \right\rangle$ approach constants, determined by the strongly coupled dynamics \cite{Affleck:1985wa}\footnote{For a calculation of the condensate in the large-$N$ limit, see \cite{Zhitnitsky:1985um}.}, in the limit $|x-y|\rightarrow \infty$. This, in turn, implies that  $\left \langle \mbox{tr} \bar\psi_+(x)\psi_-(x)\; \mbox{tr} \bar\psi_-(y)\psi_+(y) \right\rangle\sim\mbox{constant}$.\footnote{Notice that the gauging of the $U(1)$ factor is crucial for this conclusion. As the above is a finite-$N$ consideration, a nonvanishing condensate breaking a continuous global symmetry (the anomaly free chiral $U(1)$ of $2$-D QCD) in $2$-D would contradict the Coleman theorem \cite{Coleman:1973ci}.}   Therefore, from cluster decomposition, we conclude that  
\begin{equation}
\label{condensate}
\langle \tr \bar\psi_+ \psi_-  \rangle \ne 0:~~ \Z_{2N}^{d \chi} \rightarrow \Z_2~,
\end{equation}
breaking the $\Z_{2N}^{d \chi}$ discrete chiral symmetry (\ref{discretechiral}) to fermion number $\Z_2$. Similar arguments apply to the $k$-wall theory, but the bosonization rules are more involved \cite{Frishman:1992mr,Frishman:2010zz} and we simply assume (\ref{condensate}) holds. We note that $\tr  \bar\psi_+ \psi_-  $ is the only fermion bilinear which is gauge and Euclidean invariant (it equals $\tr   \psi_+ \psi_-  $ in the axial worldvolume theory of (\ref{matterkwalllagrangian})). The scenario (\ref{condensate})  with broken discrete chiral symmetry is similar to what was rigorously shown to be the case for $N=2$, where only $k$$=$$1$-walls exist  \cite{Anber:2018jdf}. 

If (\ref{condensate}) is true, the IR limit of the DW theory is ``empty" with no massless degrees of freedom. Thus, the mixed anomaly has to be matched by a TQFT describing the $N$ vacua.
Recall from (\ref{3-DCS1}) that the mixed anomaly (\ref{jacobianthooft1}) can be obtained from the variation of the $3$-D   Chern-Simons action, (\ref{3-DCS21}), which we repeat here, taking $k=1$:
\begin{equation}
 \label{3-DCS2}
 S_{3-D} =   i \;{2 \pi   \over N} \int\limits_{M_3 \;(\partial M_3 = M_2)}  {2 N A^{(1)} \over 2 \pi} \wedge
  {N B^{(2)} \over 2 \pi}   ~,
  \end{equation}
under $\delta_{\Z_{2N}} A^{(1)} = d \phi^{(0)}$, with $\phi^{(0)}\vert_{M_2} = {2 \pi \over 2 N}$ in a background $\int_{M_2}{N B^{(2)} \over 2 \pi} = p$.

A $2$-D TQFT whose quantization gives rise to  $N$ vacua  and matches the anomalous variation of  (\ref{3-DCS2})  is, see  \cite{Kapustin:2014gua}\begin{equation}\label{2-DCS}
S_{2-D} =  i \;{N  \over 2 \pi} \int\limits_{M_2}  {   \varphi^{(0)}  }  
  {  d a^{(1)}  }~.
 \end{equation}
The action (\ref{2-DCS})  has two gauge symmetries, one shifting the scalar $\varphi^{(0)}$ by $2 \pi \Z$ (this gauge symmetry can be thought to be responsible for its compactness) and the other a usual $0$-form gauge transformation of the one-form gauge field $a^{(1)}$. The gauge field $a^{(1)}$  is compact, $\oint d a^{(1)} \in 2 \pi \Z$. The gauge invariant observables are $e^{i \varphi}$ and $e^{i \oint a^{(1)}}$  and powers thereof, with correlation function (on $\R^2$) $\langle e^{i \varphi(x)} e^{i \oint_C a^{(1)}} \rangle = e^{i {2 \pi  \over N} l_{x, C}}$, with $l_{x, C}$ the linking number of $x$ and $C$ (the $N$-th powers $e^{i N \varphi}$, $e^{i N \oint a^{(1)}}$ have trivial correlation functions).

The action also has   $0$-form and   $1$-form global symmetries. 
The $\varphi^{(0)}$  compact scalar ($\oint d\varphi^{(0)} \in 2 \pi \Z$)  shifts under the $0$-form global $\Z_N$ as $ \varphi^{0} \rightarrow \varphi^{(0)} + {2 \pi \over N}$; the action remains invariant due to $a^{(1)}$ flux quantization. This scalar can be thought of as describing the phase of the fermion condensate (\ref{condensate}). 
  The $a^{(1)}$ gauge field  shifts under  $1$-form global $\Z_N^{(1)}$ as $a^{(1)} \rightarrow a^{(1)} + {1\over N}\epsilon^{(1)}$, where $\epsilon^{(1)}$ is a closed form with $\oint  \epsilon^{(1)} \in 2 \pi \Z$.
 The gauge invariant observables  $e^{i \varphi}$ and $e^{i \oint a^{(1)}}$ transform by $\Z_N$ phases under the global $0$-form and $1$-form $\Z_N$ symmetries, respectively: $e^{i \varphi} \rightarrow e^{ i {2 \pi \over N}} e^{i \varphi}$, $e^{i \oint a^{(1)}} \rightarrow e^{i {1 \over N} \oint \epsilon^{(1)}} e^{i \oint a^{(1)}}$ = $e^{i {2 \pi \Z \over N}} e^{i \oint a^{(1)}}$.

 The TQFT (\ref{2-DCS}) can be thought of as a  ``chiral lagrangian" describing the IR physics of the $N$ chiral-symmetry breaking vacua (the assumed vacua (\ref{condensate}) are gapped). This can be seen more explicitly upon
quantizing the TQFT (\ref{2-DCS}) on a finite spatial circle $\S^1$. In the temporal gauge,  $a^{(1)}_0 = 0$, one obtains the quantum mechanical action\footnote{The spatial Wilson loop of the compact $U(1)$ field $a^{(1)}$ is a compact variable, due to large gauge transformations around the $\S^1$. Gauss' law in the temporal gauge implies that $\varphi \equiv \varphi^{(0)}$ is independent of $x$. Note also that the action (\ref{qm1}) is written in Minkowski space, hence the absence of $i$.} for the compact variables $a(t) \equiv \oint\limits_{\S_1} a^{(1)}$ and $\varphi(t)$:
\begin{equation}\label{qm1}
S_{\R_t \times \S_1} = { N \over 2 \pi} \int dt \; \varphi \; {d a \over dt} 	~,
\end{equation}
  leading to the canonical commutation relations $\[ \hat\varphi, \hat a \] = - i {2 \pi \over N}$, a vanishing Hamiltonian, and the centrally extended algebra\footnote{In ref.~\cite{Anber:2018jdf}, we explicitly showed that, in  the charge-$N$ massless Schwinger model, this is the algebra of  the operators implementing discrete chiral  and center symmetry transformations. One can thus view this map as an explicit derivation of the IR TQFT from the microscopic physics.}  $e^{i \hat\varphi} e^{i \hat a} = e^{i {2 \pi \over N}} e^{i \hat a} e^{i \hat\varphi}$; as already noted,   $e^{i N \hat\varphi}$ and $e^{i N \hat a}$ are trivial operators. The Hilbert space, treating $\hat\varphi$ as coordinate, is that of $N$  states $|P\rangle$ such that $e^{i \hat\varphi} | P \rangle = | P \rangle e^{i {2 \pi P \over N}}$ and $e^{i \hat a} |P\rangle = |P + 1 ({\rm mod} N)\rangle$.
  
   The $|P\rangle$ states are the $N$ finite volume ground states due to the breaking $Z_{2N}^{d\chi} \rightarrow Z_{2}$  (\ref{condensate}), described by the expectation value of $\varphi$. 
 On the other hand, $a$, 
the spatial Wilson loop of $N$-ality one,  is an operator facilitating transitions to a neighboring vacuum.  As in the case of the Schwinger model ($N=2$) there are no physical (i.e. an intrinsic part of the gauge theory dynamics) DW in the $k$-wall theory. The role of DW on the $k$-wall worldvolume is played by   insertions of static Wilson loops  $e^{i \int_{\R_t} a^{(1)}}$, which are now defects localized in $x$, in the path integral. The correlation function $\langle e^{i \varphi(x)} e^{i \oint_C a^{(1)}} \rangle = e^{i {2 \pi  \over N} l_{x, C}}$ discussed earlier, taking a loop $C$ consisting of two infinite lines some distance apart (or, consider a compact time direction and have $C$ consist of  two Wilson loops winding in opposite directions around $\R_t$), implies that one finds neigboring vacua of the DW theory on the two sides of the  static unit $N$-ality defect.

\begin{figure}
\centering
	\includegraphics[width=4in]{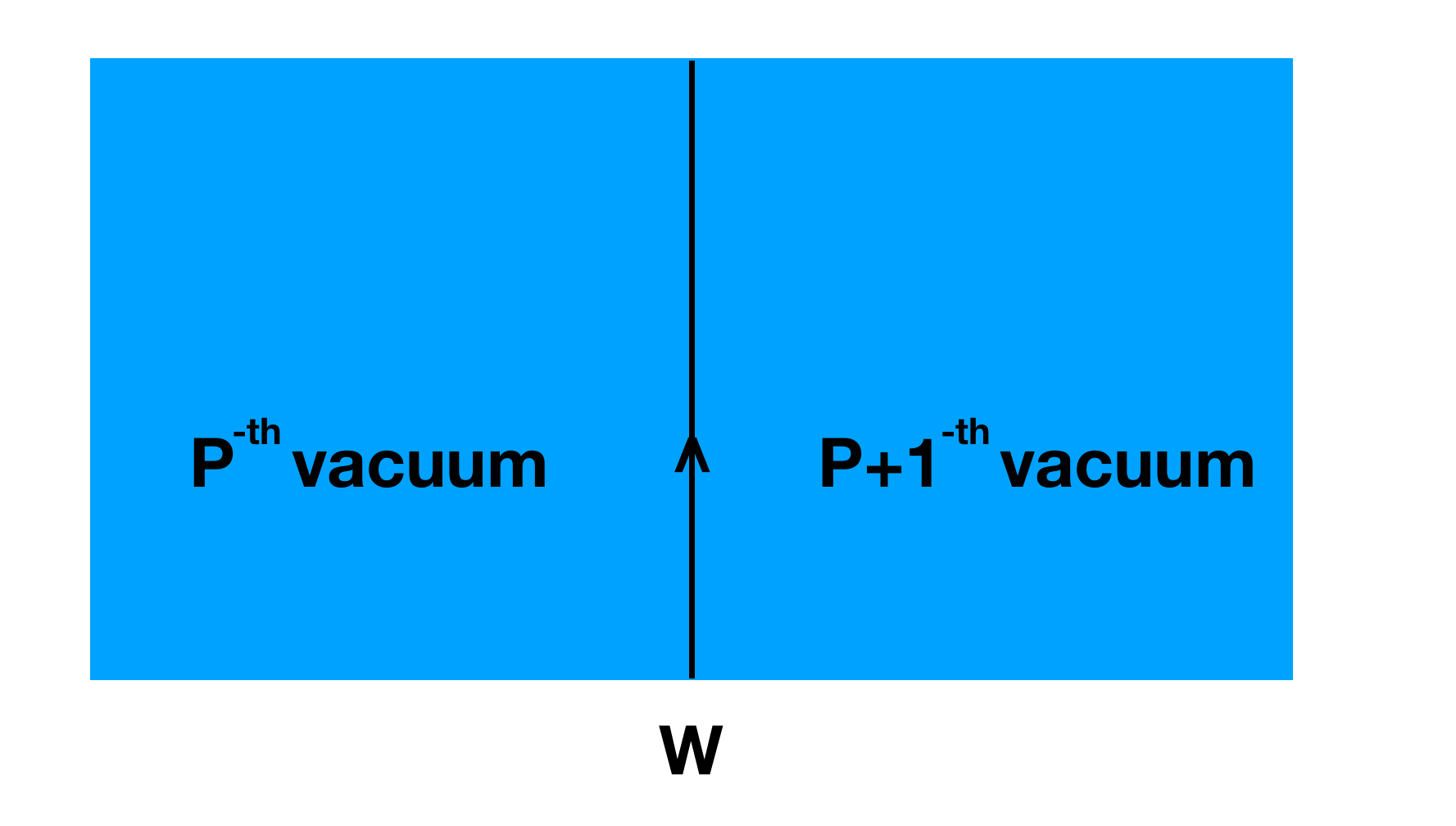}
	\caption{Two DW vacua (\ref{condensate}) separated by a fundamental quark worldline (Euclidean). As explained in Section  \ref{Screening and strings ending on walls}, $W$ can be viewed as the end of a confining string worldsheet extending into the $\R^3$ bulk. The picture holds in the high-$T$ DW on $\R^3\times S^1_\beta$, associated with center symmetry breaking. It also applies in the zero-$T$   $\R^3 \times \S^1_L$, in the semiclassically calculable $\Lambda N L \ll 1$ regime, where the DW is associated with chiral symmetry breaking. In both the small-$\beta$ and small-$L$ case, the DW worldvolume is $2$-D.
	In the small-$L$ case, the $N$ $P$-vacua are represented by distinct semiclassical  DW solutions ($N$ such solutions are known to exist for $k=1$), each carrying one-half the fundamental quark flux, see \cite{Anber:2015kea,Anber:2015wha,Poppitz:2017ivi,Anber:2017tug} for details. The resemblance between the small-$\beta$ and small-$L$ cases is because  the relevant 't Hooft anomalies on the DW are saturated in a similar mode. Note that on $\R^3\times \S^1_L$, confinement in the $\R^3$ bulk is abelian \cite{Unsal:2007jx}, in contrast to the small-$\beta$ case.}{\label{fig:01}}\end{figure}

We pause to note  that essentially the same picture---different vacua on the DW worldvolume are    separated by probe quarks---was found, by an explicit semiclassical analysis, to hold on DW between chirally broken vacua of super-YM in the calculable regime on $\R^3 \times \S^1$. While a TQFT description was not given in \cite{Anber:2015kea},  here we note that (\ref{2-DCS}) can also be used there, with  the  $0$-form $\Z_N$ of the TQFT being the $0$-form center symmetry   along the compact $\S^1$ (unbroken in the bulk, but broken on the DW). The $1$-form $\Z_N$ is the same bulk-$\R^3$ center symmetry as in the present high-$T$ discussion, see Figure \ref{fig:01} for an illustration. 

 Continuing with the high-$T$ theory, in order to see that the topological ``chiral lagrangian" (\ref{2-DCS}) matches the mixed anomaly, consider gauging the $1$-form center symmetry via the $2$-form $\Z_N$ gauge field $B^{(2)}$ (reverting back to Euclidean space and rearranging factors of $N$ and $2\pi$ in (\ref{2-DCS}) for convenience): 
\begin{equation}\label{2-DCS2}
S_{2-D} =  i \;{2 \pi  \over N} \int\limits_{M_2}  {  N \varphi^{(0)} \over 2 \pi} \wedge
  {N  (d a^{(1)} - B^{(2)}) \over 2 \pi}~,
 \end{equation}
consistent with the  gauged $1$-form invariance $a^{(1)} \rightarrow a^{(1)}+ \lambda^{(1)}$ and $B^{(2)} \rightarrow B^{(2)} + d \lambda^{(1)}$. As per our earlier discussion (see Footnote \ref{oneform}) the $1$-form transformation parameter has quantized flux $\oint d \lambda^{(1)} \in 2 \pi \Z$ and $\oint B^{(2)} = {2 \pi \Z \over N}$.\footnote{Now the $a^{(1)}$ Wilson loop observable $e^{i \oint_C a^{(1)}}$ requires a surface $\Sigma$ bounding $C$ ($C = \partial \Sigma$) in order to preserve the $1$-form gauge invariance $e^{i (\oint_C a^{(1)} - \int_{\Sigma} B^{(2)})}$. Its $N$-th power, on the other hand, is a genuine local operator,   $e^{i (N \oint_C a^{(1)} -   \oint_C B^{(1)})}$, see footnote \ref{oneform}.}  Under a chiral transformation $\delta \varphi^{(0)} = {2 \pi \over N}$, in the background of $p$ units of  't Hooft flux,  $\oint_{M_2} {N B^{(2)} \over 2 \pi} = p$, we have
 \begin{equation}\label{2-DCS21}
\delta_{\Z_N^{d \chi}} S_{2-D} =  i \;{2 \pi  \over N} \int\limits_{M_2} 
  {N  (d a^{(1)} - B^{(2)}) \over 2 \pi}~ = -i\; {2 \pi p \over N},
 \end{equation}
as required by the anomaly (\ref{jacobianthooft1}). 

Assuming that the $\Z_{2N}^{d \chi} \rightarrow \Z_2$ breaking pattern (\ref{condensate}) holds for all $k$-walls, the  TQFT describing the IR $k$-wall physics should also be given by (\ref{2-DCS}), as the  $k>1$ theory has the same number of vacua. As noted in the paragraph after (\ref{jacobianthooft1}), turning on a unit 't Hooft flux in the bulk theory corresponds to $k$ units of fractional $U(1)$ flux on the $k$-wall, i.e. $\oint_{M_2} {N B^{(2)} \over 2 \pi} = k$, so the anomaly (\ref{jacobianthooft1}) is also matched. 

\section{$\mathbf k$-walls in QCD(adj)}
\label{k-walls in QCD(adj)}
 
Finally, we comment on the $k$-walls in $SU(N)$ QCD(adj), which is a Yang-Mills theory endowed with $n_f$ adjoint Weyl fermions. As in SYM, the UV Lagrangian of this theory is invariant under a global  $U(1)_R$ axial symmetry. This symmetry, however, is anomalous  and breaks  down to the anomaly-free ${\mathbb Z^{d\chi}_{2Nn_f}}$ discrete chiral symmetry.\footnote{The breaking of $U(1)_R$ to ${\mathbb Z^{d\chi}_{2Nn_f}}$ can be easily seen from the action of $U(1)_R$ in the background of a Belavin-Polyakov-Schwarz-Tyupkin (BPST) instanton.}  In addition, the theory is invariant under a global $SU(n_f)$ symmetry, such that the adjoint fermions transform in the fundamental representation of $SU(n_f)$.\footnote{The zero-temperature theory is thought to be conformal for a range of $n_f$ ($1$$<$$n_f^*$$\le$$n_f$$<$$6$) but the precise value of $n_f^*$ is not known; see \cite{Bergner:2017gzw,Bergner:2018fxm} and \cite{Anber:2018tcj,Cordova:2018acb,Bi:2018xvr} for   recent lattice results and theoretical discussions, respectively.}

 Everything we said about the wall action in SYM transcends naturally to QCD(adj); the only difference is an additional factor of $n_f$ multiplying  $(-1)^n$  in (\ref{k-wall action}), which amounts to scaling the wall tension by a trivial numerical coefficient. The worldvolume of the $k$-wall is also a  $2$-D QCD with gauge group $U(1)\times SU(N-k)\times SU(k)$ and fermions charged under $U(1)$ and transforming in the bi-fundamental representation of $SU(N-k)\times SU(k)$. In addition, as in the UV theory,  the fermions transform in the fundamental representation of the global $SU(n_f)$.\footnote{As in the $N=2$ case  \cite{Anber:2018jdf}, for $n_f>1$ four-fermion terms on the $k$-wall worldvolume reduce the $SU(n_f)_+ \times SU(n_f)_-$ chiral symmetry of the kinetic terms of the worldvolume theory to the diagonal $SU(n_f)$ of the bulk.} Under an axial $U(1)_R$ transformation $\psi_{\pm}\rightarrow e^{i\chi}\psi_{\pm}$ the measure transforms as in (\ref{anomaly$2$-D0}), with ${\cal J}$ now replaced by
\begin{equation}
 {\cal J} \equiv \exp\left[ i \; 2 n_f \chi (N-k)k \; \gamma^{(N-k)} \oint {F_{12}^{N-k} dx^1 dx^2   \over 2\pi} \right]\,,
\end{equation}
i.e., there is an extra factor of $n_f$ in the Jacobian. Repeating the same steps from (\ref{anomaly$2$-D0}) to (\ref{jacobian1}), one can easily see that there is an anomaly-free $\mathbb Z^{d\chi}_{2n_fN}$ discrete chiral symmetry on the DW. Similarly, it is straightforward to see that  there is a  mixed discrete 't Hooft anomaly upon turning on a $p$-twist of $SU(N)$: ${\cal J}=e^{- i {2\pi \over N} k p}$. 

Yet, the most interesting part of the story is the fate of the discrete $\mathbb Z_{2n_fN}^{d\chi}$ and global $SU(n_f)$ symmetries on the wall. Since our theory lives in $2$-D, one expects that $SU(n_f)$ remains unbroken in the IR, as suggested by the Coleman theorem \cite{Coleman:1973ci}. Interestingly, one can use the nonabelian bosonization and WZW model,  discussed in Section \ref{section:chiralanomaly}, to show that this is indeed the case. Let us for simplicity consider the $k=1$-wall. Now, since there is an extra $SU(n_f)$ global symmetry, the bosonization rule (\ref{bosonization rule}) should be replaced by\footnote{Again, as in Section \ref{section:chiralanomaly}, the correlators here refer to the vectorlike version of the axial worldvolume theory, obtained by relabeling $\psi_+ \leftrightarrow \bar\psi_+$. }
\begin{eqnarray}
\bar\psi_+^{a\,i}\psi_{-b\,j}=\mu\, h^{a}_b\; g^{i}_j \; e^{-i\sqrt{\frac{4\pi}{n_f(N-1)}}\phi}\,,
\label{bosonization rule general}
\end{eqnarray}
where, as before, $h$ and $e^{-i\sqrt{\frac{4\pi}{n_f(N-1)}}\phi}$ are  boson fields, $SU(N-1)$- and $U(1)$-group valued, respectively, while $g$ is a  global-$SU(n_f)$ group-valued  boson-field matrix. 
There are well-known subtleties with the above multi-flavor bosonization rule, which, however, have been argued to be not important for studying the low-energy physics in the strong coupling   limit $e \gg m \rightarrow 0$ ($e$ is the $2$-D gauge coupling and $m$ the fermion mass) \cite{Affleck:1985wa}, \cite{Frishman:1992mr}. This  is also the limit considered here and, for our qualitative considerations, we shall assume that  (\ref{bosonization rule general}) holds.
As before,  the correlators $ \langle e^{-i\sqrt{\frac{4\pi}{n_f(N-1)}}\phi(x)}e^{ i\sqrt{\frac{4\pi}{n_f(N-1)}}\phi(y)}\rangle $ and $\left\langle \mbox{tr}  h^\dagger(x) \;\mbox{tr} h(y) \right\rangle$ approach constants in the limit $|x-y|\rightarrow \infty$, thanks to the gauging of $U(1)\times SU(N-1)$. The correlator $\langle g_{i}^j(x)g_{k}^{l}(y) \rangle$, however, behaves as \cite{Knizhnik:1984nr}
\begin{eqnarray}
\left\langle g_{i}^j(x)g_{k}^{l}(y)\right\rangle=\frac{\delta_{i}^l\delta_{k}^j}{\left[M|x-y| \right]^{\frac{n_f^2-1}{n_f(n_f + N-1)}}}\,,
\label{g correlator}
\end{eqnarray}
and $M$ is a mass scale. Next, we define the $SU(N-1)$ color-singlet operator 
\begin{eqnarray}\label{o1}
{\cal O}^{(1)\,i}_j\equiv \bar\psi_+^{a\,i}\psi_{-a\,j}\,,
\end{eqnarray}
which transforms non-trivially under $SU(n_f)$. Then, we can use (\ref{bosonization rule general}) and (\ref{g correlator}) to show that $\langle {\cal O}^{(1)\,\dagger i}_j(x) {\cal O}^{(1)\,i}_j(y) \rangle\rightarrow 0$ as  $|x-y|\rightarrow \infty$. Therefore, we find
\begin{eqnarray}
\langle \bar\psi_+^{a\,i}\psi_{-a\,j} \rangle=0\,,
\end{eqnarray}
and conclude that $SU(n_f)$ is unbroken in the IR, in accord with the Coleman theorem. 

What remains is to examine the discrete chiral symmetry $\mathbb Z_{2n_fN}^{d\chi}$. To this end we consider the color-singlet and $SU(n_f)$-singlet operator
\begin{eqnarray}
{\cal O}^{(2)}(x) \equiv  \det\limits_{i,j} \bar\psi_+^{a \,i }(x) \psi_{-a \,j }(x)  .
\label{O2 operator}
\end{eqnarray}
It is trivial to see that ${\cal O}^{(2)}$ acquires a phase $e^{i\frac{2\pi}{N}}$ under a $\mathbb Z_{2n_fN}^{d\chi}$ transformation, and hence, it can be used to examine the breaking of $\mathbb Z_{2n_fN}^{d\chi}$. As it is an $SU(n_f)$ singlet, it is possible that  the correlator (\ref{g correlator}) of the $SU(n_f)$-valued bosonic field $g^i_j$, which disorders the fermion bilinear (\ref{o1}), does not similarly affect the ${\cal O}^{(2)}$ two-point correlation function. If so,  $\langle {\cal O}^{(2)}(x)  {\cal O}^{(2) \;\dagger}(y)\rangle$ would approach constant at infinite $|x-y|$. Then,  $\mathbb Z_{2n_fN}^{d\chi}$ is broken down to $\mathbb Z_{2n_f}^{d\chi}$ on the $k=1$-wall leading to $N$ distinct vacua on the wall, a  result that also generalizes to the $k$-wall.  In this scenario, the IR spectrum of the  $k$-wall in QCD(adj) would be free from massless excitations, and the mixed anomaly would be matched by a TQFT describing the $N$ vacua, exactly as in SYM.

{\flushleft \bf Acknowledgments:} We thank Zohar Komargodski for discussions. MA is supported by an NSF grant PHY-1720135. EP is supported by a Discovery Grant from NSERC.

 \appendix 

\section{Group theory conventions} 
\label{appendix:groups}
 
We denote the fundamental $SU(N)$ generators by $H^a$, $a=1,...,N-1$. An explicit form is 
$H^a = {\rm diag}[\lambda^{a1},..., \lambda^{aN}]$, where 
\begin{eqnarray}
\label{group1}
H^a &=& {\rm diag}[\lambda^{a1},..., \lambda^{aN}] ={1 \over \sqrt{a(a+1)}} {\rm diag}[\underbrace{1,1,...,1}_{a\,\mbox{\small times}},-a,\underbrace{0,0,...0}_{N-1-a \, \mbox{\small times}}]  ~\\
\lambda^{a A} &\equiv& {1 \over \sqrt{a(a+1)}}(\theta^{aA} - a \delta_{a+1,A}) ,  ~a = 1,...,N-1,~A=1,...,N,~\theta^{aA} \equiv \left\{\begin{array}{c} 1, a \ge A\cr 0, a < A\end{array} \right. \nonumber
\end{eqnarray}
 The only utility in introducing $\lambda^{aA}$ in (\ref{group1}) is  to note that the weights of the fundamental representation $\bm\nu^A$ can be expressed in this $N-1$-dimensional basis (we denote its $a$-th component by $(\bm\nu^A)_a$) as:
  \begin{eqnarray}  
\label{weightsoffund}  (\bm\nu^A)_a = \lambda^{aA},~ ~ \bm\nu^A \cdot \bm\nu^B \equiv \sum\limits_{a=1}^{N-1} \lambda^{a A} \lambda^{a B} = \delta^{AB} - {1 \over N}, ~ \sum\limits_{A=1}^{N} \lambda^{a A} \lambda^{b A} = \delta^{ab}~, 
  \end{eqnarray}
  where we also noted the properties of the $\lambda^{aA}$ implying that $\tr H^a H^b = \delta^{ab}$. 
 
 Furthermore, the fundamental weights and simple roots of $SU(N)$ which we denote by $\bm \omega^a$ and $\bm\alpha^a$, respectively,   are 
 \begin{eqnarray}
 \label{root weights}
 \bm \omega^a &=& \sum\limits_{A=1}^a \bm \nu^A, ~ a = 1, \ldots N-1,\\
 \bm \alpha^a &=&  \bm \nu^a - \bm\nu^{a+1}, ~ a = 1, \ldots N-1. \nonumber
  \end{eqnarray}
We also define the positive roots $\bm \beta^{AB}$, $A<B$:
  \begin{equation}
  \label{betaroots}
  \bm \beta^{AB} \equiv \bm \nu^A  - \bm \nu^{B}~, ~ A, B = 1, \ldots N.   \end{equation}
The simple roots are a subset, $\bm \alpha^a = \bm \beta^{a \; a+1}$ and  the affine root is $\bm \alpha^0 = - \sum\limits_{k=1}^{N-1} \bm\alpha^k$. 
We shall need several relations that  follow from the  definitions (\ref{group1}, \ref{weightsoffund}, \ref{root weights}):
 \begin{eqnarray}
 \label{relation1}
 \bm \omega^a \cdot \bm \omega^b &=& {\rm min}(a,b) - {a b \over N} \\
 \label{relation2}
 \bm \omega^b \cdot \bm H &=& \sum\limits_{a=1}^{N-1} (\bm \omega^b)_a H^a = {\rm diag}[\underbrace{1 -{b \over N},1 -{b \over N},...,1 -{b \over N}}_{b\,\mbox{\small times}},\underbrace{-{b \over N},-{b \over N},...-{b \over N}}_{N-b \, \mbox{\small times}}] \\
 \label{relation3}
 \bm \omega^b \cdot \bm \beta^{AB} &=& \sum\limits_{k=1}^b  \delta^{kA}-\delta^{kB}  =  \left\{\begin{array}{cc}0,  & b <A \cr 1,  &  A \le b < B   \cr 0,  & b \ge B \end{array}\right. ~.
 \end{eqnarray}

Next, as will be seen below,  on the DW the $SU(N)$ group breaks to $U(1)\times SU(k)\times SU(N-k)$. Here we introduce some algebraic notation that will be useful to study the DW theory. We define the unbroken $U(1)$ generator as 
\begin{eqnarray}
\label{tildeH}
\tilde{H}^{N-k}=\frac{1}{\sqrt{kN(N-k)}}\mbox{diag}\left[\underbrace{k,k,...,k}_{N-k\,\mbox{\small times}},\underbrace{k-N,k-N,...,k-N}_{k\,\mbox{\small times}}\right]\,,
\end{eqnarray}
 satisfying $\mbox{tr}\left[\tilde{H}^{N-k}\tilde{H}^{N-k}\right]=1$ (we use the tilde to stress  that this is not one of the $H^a$ previously introduced in (\ref{group1}), but can be expressed as their linear combination). 
Further, we break the Lie-algebra generators of $SU(N)$ as follows (omitting $\tilde{H}^{N-k}$ (\ref{tildeH}), which commutes with the $SU(N-k)$ and $SU(k)$ hermitean generators  ${\cal T}^a$ and ${\cal T}^A$)
\begin{eqnarray}\label{kwallgenerators}
T=\left[\begin{array}{cc} {\cal T}^a_{(N-k)\times(N-k)} & E_{\bm\beta\,(N-k)\times k}\\ E_{-\bm\beta\,k\times(N-k)} & {\cal T}^A_{k\times k}  \end{array} \right]\,,
\end{eqnarray}
where the subscript indicates the matrix dimensionality. 
There are $a=1,2,..,(N-k)^2-1$ $SU(N-k)$ generators ${\cal T}^a$, $A=1,2,...,k^2-1$ $SU(k)$ generators ${\cal T}^A$, and $2k(N-k)$ generators $E_{\pm \bm\beta}$ corresponding to $2k(N-k)$ different roots of $SU(N)$. This adds up to the original $N^2-1$ generators of $SU(N)$. We shall not need to explicitly define the ${\cal T}^a$ and ${\cal T}^A$ generators.

Explicitly, we define the $(N-k)\times k$ matrix $E_{\bm\beta\,}$ as follows. First, to enumerate the roots $\beta$, we denote them as $\beta_{AA'}$, with $A=1,...,N-k$ and $A'=1,...,K$. Next, regarding $E_{\bm\beta_{AA'}}$ as an $(N-k)\times k$ matrix, we define its matrix elements as 
\begin{equation}
(E_{\bm\beta_{AA'}})_{BB'} = \delta_{AB} \delta_{A'B'}, ~ B=1,...,N-k, ~~ B'=1,...,k . 
\label{Ebeta}
\end{equation}
Clearly, these are $k(N-k)$ linearly independent generators. The matrices $E_{-\bm\beta}$ are similarly defined: these are $k \times (N-k)$ matrices, labeled by $-\bm\beta_{AA'}$, with matrix elements 
\begin{equation}
(E_{-\bm\beta_{AA'}})_{B'B} = \delta_{AB} \delta_{A'B'}, ~ B=1,...,N-k, ~~ B'=1,...,k . 
\label{Eminusbeta}
\end{equation} In matrix form, we have that $E_{-\bm\beta_{AA'}} = (E_{\bm \beta_{AA'}})^T$.\footnote{We note that the roots $\bm\beta_{AA'}$ just defined should not be confused with  $\bm\beta^{AB}$ of (\ref{betaroots}) (they can be related but it is notationally challenging). We avoid introducing special notation to distinguish them, other than denoting them with lower-subscript indices.}
The explicit form (\ref{Ebeta}, \ref{Eminusbeta}) of $E_{\pm \bm \beta}$ implies the following relations that we shall use in the following:
\begin{eqnarray}
 \tr (E_{\bm \beta_{CC'}} E_{- \bm \beta_{BB'}})&=&\delta_{CB} \delta_{C'B'} \nonumber \\
 \tr(E_{\bm \beta_{CC'}} {\cal T}^A E_{- \bm \beta_{BB'}})&=& \delta_{CB} {\cal T}^A_{C'B'} \\
 \tr(E_{- \bm\beta_{CC'}} {\cal T}^a E_{\bm \beta_{BB'}} ) &=& \delta_{C'B'} {\cal T}^a_{CB} \nonumber. 
\end{eqnarray}

It is now straightforward to prove the relations that follow (the index convention below is the one from the previous paragraph: primed indices range from $1,...,k$ and unprimed from $1,...,N-k$):
\begin{eqnarray} \label{HE}
\left[\tilde{H}^{N-k},E_{\pm\bm \beta_{AA'}}\right]&=&\pm \gamma^{(N-k)}E_{\pm\bm\beta_{AA'}}\,, ~~ \gamma^{(N-k)} \equiv \sqrt{\frac{N}{k(N-k)}}, \nonumber \\
 (\left [{\cal T}^{a}, E_{\bm \beta_{AA'}}\right])_{BB'}&\equiv & 
({\cal T}^{a}  E_{\bm \beta_{AA'}})_{BB'}  = \sum_{C} ({\cal T}^{a})_{BC} (E_{\bm\gamma_{AA'}})_{CB'}\,, \nonumber \\
(\left [{\cal T}^{a}, E_{-\bm \beta_{AA'}}\right])_{B'B}&=& - (E_{-\bm \beta_{AA'}} {\cal T}^{a})_{B'B} = -\sum_{C}   (E_{-\bm\beta_{AA'}})_{B'C} ({\cal T}^{a})_{CB}\,,\\
 (\left [{\cal T}^{A}, E_{\bm \beta_{AA'}}\right])_{BB'}&=& -(E_{\bm \beta_{AA'}}{\cal T}^{A})_{BB'}=-\sum_{C'}   (E_{\bm \beta_{AA'}})_{BC'} ({\cal T}^{A})_{C'B'}\,,\quad \nonumber 
 \\ 
 (\left [{\cal T}^{A}, E_{-\bm \beta}\right])_{B'B} &=& ({\cal T}^{A} E_{-\bm \beta})_{B'B} 
 =
 \sum_{C'}  ({\cal T}^{A})_{B'C'} (E_{-\bm\beta_{AA'}})_{C'B}\,. \nonumber
\end{eqnarray}
We kept the top equation in (\ref{HE}) in matrix form, where we extended the $(N-k)\times k$ matrix $E_{ \bm\beta}$  to an $N\times N$ matrix by embedding  as in (\ref{kwallgenerators}). The rest of the equations representing the action of the $SU(N-k)$ and $SU(k)$ generators was shown explicitly using index notation, consistent with the definitions (\ref{Ebeta}, \ref{Eminusbeta}, \ref{kwallgenerators}).
  
   \section{Fermion zero modes}
\label{appendix:zeromodes}
To find the fermion zero modes, we begin with 
the  covariant derivative containing the $\lambda^\beta$ fermions, as explained in Section \ref{section:fermions} of the main text. Using the decompositions (\ref{kwalldecompositions}), and recalling the action of ${\cal T}^{a, A}$ on $E_{\pm \beta}$ of (\ref{HE}) which we use below in matrix form (a summation over $C$ is from $1,..., N-k$ and $C'$ from $1,...,k$ is understood below), it is given by
\begin{eqnarray} \label{kwalllambda1}
\partial_\mu \lambda - i [A_\mu, \lambda] &=&  ~~ \partial_\mu \lambda^{\bm \beta_{CC'}} E_{\bm \beta_{CC'}} +  \partial_\mu \lambda^{-\bm \beta_{CC'}} E_{-\bm \beta_{CC'}} \nonumber \\
&& - i \gamma^{(N-k)} A_\mu^{N-k} (\lambda^{\bm \beta_{CC'}} E_{\bm \beta_{CC'}} -  \lambda^{-\bm \beta_{CC'}} E_{-\bm \beta_{CC'}})  \nonumber 
\\
&&- i A^a_\mu \;  ( {\cal T}^a   E_{\bm\beta_{CC'}}  \lambda^{\bm \beta_{CC'}} - E_{-\bm\beta_{CC'}} {\cal T}^a \lambda^{-\bm \beta_{CC'}}) \nonumber \\
&&+ i A^A_\mu \;  ( E_{\bm \beta_{CC'}} {\cal T}^A \lambda^{\bm \beta_{CC'}}  -{\cal T}^A E_{- \bm \beta_{CC'}} \lambda^{- \bm \beta_{CC'}}) ~
\end{eqnarray}
Because the matrices $E_{\bm \beta}$, ${\cal T}^{a}E_{\bm \beta}$, $E_{\bm \beta}{\cal T}^{A}$ and the matrices $E_{-\bm \beta}$, ${\cal T}^{A} E_{-\bm \beta}$, $E_{-\bm \beta}{\cal T}^{a}$ are in orthogonal subspaces of $SU(N)$,   we can write the equations of motion for $\lambda^{\bm \beta}$ and $\lambda^{-\bm\beta}$ separately as follows:
\begin{eqnarray}
\label{kwalllambda2}
  \bar\sigma^\mu &  (& \partial_\mu  \lambda^{\bm \beta_{CC'}} E_{\bm \beta_{CC'}}  - i \gamma^{(N-k)} A_\mu^{N-k}    \lambda^{\bm \beta_{CC'}} E_{\bm \beta_{CC'}}  \nonumber \\ &&  - i A^a_\mu \;    {\cal T}^a   E_{\bm\beta_{CC'}}  \lambda^{\bm \beta_{CC'}}+ i A^A_\mu \;  E_{\bm \beta_{CC'}} {\cal T}^A \lambda^{\bm \beta_{CC'}} ) = 0 ~,\nonumber \\
 \bar\sigma^\mu  & (& \partial_\mu  \lambda^{-\bm \beta_{CC'}} E_{-\bm \beta_{CC'}}  + i \gamma^{(N-k)} A_\mu^{N-k}    \lambda^{-\bm \beta_{CC'}} E_{-\bm \beta_{CC'}}   \nonumber \\ &&+ i A^a_\mu \;   E_{-\bm\beta_{CC'}}   {\cal T}^a   \lambda^{-\bm \beta_{CC'}}- i A^A_\mu \;  {\cal T}^A E_{-\bm \beta_{CC'}}  \lambda^{-\bm \beta_{CC'}} ) = 0 ~.
\end{eqnarray}
We now multiply the first equation in (\ref{kwalllambda2})  by $E_{- \bm\beta_{BB'}}$ and take the trace to obtain the equation   for $\lambda^{\bm \beta_{BB'}}$:
\begin{eqnarray}
\label{lambdabetaeqn}
0&=&\bar\sigma^\mu \left( \partial_\mu \lambda^{\bm \beta_{BB'}} - i A^{N-k}_\mu \gamma^{(N-k)}  \lambda^{\bm \beta_{BB'}} - i A_\mu^a {\cal T}^a_{BD} \lambda^{\bm \beta_{D B'}} + i A_\mu^A \lambda^{\bm \beta_{BD'}}  {\cal T}^A_{D'B'}\right)  .
\end{eqnarray}
Eqn. (\ref{lambdabetaeqn}) shows that  $\lambda^{\bm \beta_{BB'}} $, considered as the $BB'$-th element of a  $(N-k) \times k$ matrix $\lambda^{\bm \beta}$, transforms as   $\lambda^{\bm \beta} \rightarrow U_{N-k} \; \lambda^{\bm \beta} \; U^\dagger_k$  under $SU(N-k) \times SU(k)$ gauge transformations, where $U_{N-k} = e^{ i \omega^a {\cal T}^a}$, $U_k = e^{i \omega^A {\cal T}^A}$. It is easy to see that the (\ref{lambdabetaeqn}) is invariant under these transformation, along with $A_\mu^a {\cal T}^a \rightarrow U_{N-k} (A_\mu^a {\cal T}^a + i \partial_\mu) U_{N-k}^\dagger$ and $A_\mu^A {\cal T}^A \rightarrow U_{k} (A_\mu^A {\cal T}^A + i \partial_\mu) U_{k}^\dagger$.
 
Next, proceeding as in the derivation of  (\ref{kwalllambda2}), multiplying the second equation by $E_{\bm \beta_{BB'}}$ and taking the trace, we obtain the equation of motion for $\lambda^{-\bm\beta_{BB'}}$:
 \begin{eqnarray}
 \label{lambdaminusbetaeqn}
 0&=&\bar\sigma^\mu \left( \partial_\mu \lambda^{-\bm \beta_{BB'}} + i A^{N-k}_\mu \gamma^{(N-k)}  \lambda^{-\bm \beta_{BB'}} + i  \lambda^{-\bm \beta_{D B'}} {\cal T}^a_{DB} A_\mu^a - i A_\mu^A {\cal T}^A_{B'D'} \lambda^{-\bm \beta_{BD'}} \right) 
 \end{eqnarray}
 Considering $\lambda^{-\bm\beta_{BB'}}$ as the $B'B$-th entry of a $k \times (N-k)$ matrix $\lambda^{-\bm\beta}$, we conclude that it transforms as $\lambda^{-\bm\beta} \rightarrow U_k \; \lambda^{-\bm\beta} U_{N-k}^\dagger$ under $SU(N-k)\times SU(k)$ gauge transformations.
 
Finally,  the transformation of $\lambda^{\bm \beta}$, $\lambda^{-\bm \beta}$ under the $U(1)$ as $\lambda^{\pm \beta} \rightarrow e^{\pm i \omega \gamma^{(N-k)}} \lambda^{\pm \beta}$, with $A_\mu^{N-k} \rightarrow e^{i \omega}(A_\mu^{N-k} + i \partial_\mu)e^{-i \omega}$; these are inherited from the $SU(N)$ transformation $e^{i \omega \tilde{H}^{N-k}}$.
 The transformation properties of the fermions under the unbroken gauge group on the $k$-wall are summarized in 
 Table \ref{charges of k DW}.

 To find the fermion zero modes in the $k$-wall background, 
we now restrict the gauge backgrounds in the equations of motion (\ref{lambdabetaeqn},\ref{lambdaminusbetaeqn}) to the $k$-wall background (\ref{kwallsolution}). We also decompose the 
fermions into Matsubara modes, $\lambda^{\pm \beta}_p$, defined as $\lambda^{\pm\beta}(x^4, x^i) = \sum_{p \in \Z} \lambda_p^{\pm\beta}(x^i) e^{i 2 \pi T p' x^4}$, where $p'\equiv p+1/2$. In the $k$-wall background, the equations of motion for the different Matsubara modes decouple. 

The $k$-wall is orthogonal to the $z=x^3$ direction, with worldvolume along $x^{1,2}$ and, recalling that $\bar\sigma^\mu = (\bm\sigma,-i \sigma^0)$, we obtain the $z$-dependent Weyl equation
\begin{equation}
\label{weylkwall}
0= \left[ \sigma^3 \partial_z  + \sigma^0 \left(2 \pi p' T \mp TQ^{(k)}(z) \gamma^{(N-k)} \right) \right] \lambda^{\pm \bm\beta}_p ~,
\end{equation}
 where the signs are correlated. The solution of this equation  is   \begin{equation}
\label{zeromodesolution} 
\lambda_p^{\pm\bm\beta}(z) = e^{~\sigma^3 \left[ - 2\pi p' T z \pm T \gamma^{(N-k)} \int\limits_0^z Q^{(k)}(z') dz'\right]} \; \lambda_p^{\pm\bm\beta}(0).
 \end{equation}
 We now recall from (\ref{qbc}) that  $Q^{(k)}(z)$ vanishes as $z\rightarrow -\infty$ and $\int\limits^{z \rightarrow - \infty}_0 Q^{(k)}(z') dz'$ converges. Thus normalizability of the solutions as $z\rightarrow -\infty$ is only determined by the first term in the exponent of (\ref{zeromodesolution}),  requiring that only  wave functions $\lambda_p^{\pm\bm\beta}(0)$ which are eigenstates of $\sigma^3$ with  $\sigma^3 p' <0$   lead  to a zero mode  normalizable as $z\rightarrow -\infty$. On the other hand, as $z \rightarrow \infty$, the boundary conditions imply that $\gamma^{(N-k)} \int\limits^{z\rightarrow + \infty}_0 Q^{(k)}(z') dz' \rightarrow - 2\pi z $. Thus normalizability of (\ref{zeromodesolution}) at $z\rightarrow + \infty$ requires that $\lambda_p^{\pm\bm\beta}(0)$  also  be eigenvalues of $\sigma^3$ with $- \sigma^3 p' \mp \sigma^3 <0$. 
 
 To proceed, we recall that $\lambda^{\pm\beta}_p$ are two-component Weyl spinors. 
 We denote their upper components $\lambda^{\pm\beta}_{p,1}$ (with $+1$ eigenvalue of $\sigma^3)$ and the lower components by $\lambda^{\pm\beta}_{p,2}$ (with $-1$ eigenvalue of $\sigma^3)$. We further note that the upper $\lambda^{\pm\beta}_{p,1}$ (lower  $\lambda^{\pm\beta}_{p,2}$) components obey a positive (negative) two-dimensional  chirality  Weyl equation, as follows from our definition of $\bar\sigma^\mu$.\footnote{The $2$-D positive chirality, or left mover, Euclidean Weyl equation is $(\partial_1 + i \partial_2 )\psi_+ = 0$ (the negative chirality, or right mover, one is $(\partial_1 - i \partial_2) \psi_- = 0$).}
 The  two conditions for normalizability stated after (\ref{zeromodesolution}) admit (recalling $p' = p+1/2$) only the following solutions:
 \begin{eqnarray}
 \label{kwallzeromodes}
 \lambda^{\bm\beta}_{p=-1, 1}:&~& \psi_+,\;\; {\rm left \; mover}, ( \gamma^{(N-k)}, \Box, \overline\Box )\; {\rm under} \;(U(1), SU(N-k), SU(k)), \nonumber \\
  \lambda^{-\bm\beta}_{p=0, 2}:&~& \psi_-,\;\;  {\rm right \; mover}, (- \gamma^{(N-k)}, \overline\Box, \Box)\; {\rm under} \;(U(1), SU(N-k), SU(k)),
 \end{eqnarray}
 where we also display their quantum numbers under the unbroken gauge group on the $k$-wall.
 The information contained in (\ref{kwallzeromodes}) is also shown in Table \ref{charges of k DW}. There, we also note that the zero modes have the same charge under the anomalous chiral symmetry of the bulk $4$-D theory as they originate in the same $4$-D Weyl fermion. 
Finally, as in the case of the $SU(2)$ theory of ref.~\cite{Anber:2018jdf},  the $2$-D $k$-wall worldvolume theory is an axial one: the $R-$ and $L$-moving fermions have opposite charges under all gauge groups; nonetheless, it is clear that the $2$-D $U(1)\times SU(N-k)\times SU(k)$ gauge theory with matter content given in (\ref{kwallzeromodes}) is gauge anomaly free. For brevity, we rename the two $2$-D Weyl fermions $\psi_+$ and $\psi_-$ as shown also in  Table \ref{charges of k DW}.
 
 \begin{table}[h]
\begin{center}
\begin{tabular}{|c|c|c|c|c|c}
\hline
&  $\lambda^{\bm \beta}$ & $\lambda^{-\bm \beta}$\\\hline
gauge $U(1)$ & $\gamma^{(N-k)} = \sqrt{\frac{N}{k(N-k)}}$ & $-\gamma^{(N-k)}=-\sqrt{\frac{N}{k(N-k)}}$\\\hline
gauge $SU(k)$ & $\overline\Box$ & $\Box$\\\hline
 gauge $SU(N-k)$ & $\Box$ & $\overline\Box$\\ \hline\hline
 $2$-D chirality  & left mover & right mover\\\hline
global $U(1)_{R}$ & 1 & 1\\\hline
 $2$-D field & $\psi_+$ &$\psi_-$ \\\hline
\end{tabular}
\caption{ \label{charges of k DW} Charges of the fermions $\lambda^\beta$ (see (\ref{kwalldecompositions})) under the unbroken gauge group  on the $k$-wall. In the bottom two rows, we show the normalizable zero modes' $2$-D chirality and their charge under the anomalous $4$-D $U(1)_R$ chiral symmetry. The bottom line in the table introduces the $2$-D notation for the Weyl fermions of the $k$-wall theory, see also eq.~(\ref{kwallzeromodes}).}
\end{center}
\end{table}
%

 \section{$\mathbf{U(1)}$ flux  quantization}
 \label{appendix:quantization}

 In order to understand $U(1)$ charge quantization, we need to recall the description of boundary conditions on the torus as in \cite{tHooft:1979rtg,vanBaal:1982ag}. We begin by recalling the description of gauge bundles in the $U(1) \times SU(N-k) \times SU(k)$ theory on the two-torus ($0 \le x_1 \le L_1$, $0 \le x_2 \le L_2$). These are described by transition functions $\Omega_{i \; [1]}, \Omega_{i \; [N-k]}, \Omega_{i \; [k]}$, where $[1], [N-k], [k]$ indicates the gauge group and $i=1,2$ the direction on the torus. The gauge fields obey the boundary conditions:
 \begin{eqnarray}\label{boundary}
 A_{\;[b]}(L_1,x_2) &=& \Omega_{1 \;[b]}(x_2) \left(A_{\; [b]}(0,x_2) + i d \right) \Omega^\dagger_{1 \; [b]} (x_2)~, \nonumber \\
  A_{\;[b]}(x_1,L_2) &=& \Omega_{2 \;[b]}(x_1) \left(A_{\; [b]}(x_1,0) + i d \right) \Omega^\dagger_{2 \; [b]} (x_1)~,
 \end{eqnarray}  where  ~{\rm $ [b]=[1],[N-k],[k]$. 
The transition functions $\Omega_{i \;[b]}$ are gauge group elements (the $U(1)$ boundary conditions are also given in (\ref{boundaryu1}) below). For $SU(N-k)$ and $SU(k)$, they are   group elements in the defining representation. For later use, we embed  them in $SU(N)$ as in (\ref{kwallgenerators}):
 \begin{eqnarray}
\label{sunktransitionfunctions}
\hat\Omega_{i \,[N-k]} =\left[\begin{array}{cc} \Omega_{i \,[N-k]} &0 \\0&   I_{k}  \end{array}\right] ~,
\end{eqnarray}
where we indicated by $\hat\Omega$ that the transition function is embedded into $SU(N)$ and $I_k$ is a $k\times k$ unit matrix (in the following, we shall often omit the hat and the context will show whether   the above embedding is used). 
Similarly, for $SU(k)$, we have 
 \begin{eqnarray}
\label{suktransitionfunctions}
\hat\Omega_{i \,[k]} =\left[\begin{array}{cc} I_{N-k}&0 \\0&   \Omega_{i \,[k]}   \end{array}\right] ~,
\end{eqnarray}
while the transition functions for $U(1)$ follow from the form of its generator, eqn.~(\ref{tildeH})
 \begin{eqnarray}
\label{u1transitionfunctions}
\Omega_{i \,[1]} =\left[\begin{array}{cc} e^{i2\pi \omega_{i} \frac{k}{N} \gamma^{(N-k)}} I_{N-k} &0 \\0&  e^{i2\pi \omega_{i} \frac{k -N}{N} \gamma^{(N-k)}} I_{k}  \end{array}\right] ~.
\end{eqnarray}

Further,  from (\ref{boundary}) it follows that under gauge transformations $g_{[b]}(x_1, x_2)$ ($0 \le x_1 \le L_1$, $0 \le x_2 \le L_2$)
  the transition functions  transform  \begin{eqnarray}
\label{transitiontransform}
A_{\; [b]} &\rightarrow& g_{[b]} (A_{\; [b]} + i d) g_{[b]}^\dagger \nonumber  \\
\Omega_{1 \; [b]}(x_2)  &\rightarrow &g_{[b]}(L_1, x_2) \; \Omega_{1 \; [b]}(x_2)  \; g^\dagger_{[b]}(0, x_2)~,  \\
\Omega_{2 \; [b]}(x_1) &\rightarrow &g_{[b]}(x_1, L_2) \; \Omega_{2 \; [b]} (x_1) \; g^\dagger_{[b]}(x_1, 0)~.\nonumber
\end{eqnarray}
The above gauge transformations and boundary conditions imply    also  that  a fundamental Wilson loop winding the $i$-th direction of the torus requires an insertion of a transition function in order to be gauge invariant. For example,  for $i=1$,
\begin{equation}\label{wilsonloop}
W_{[b]} (x_2) = \tr( e^{i \int\limits_{ 0}^{L_1} A_{1 \; [b]} (x_1,x_2) dx^1} \Omega_{1 \; [b]} (x_2) )~.
\end{equation}

The transition functions must 
 further  obey the consistency conditions known as cocycle conditions,  following from requiring that $A_{\; [b]} (L_1,L_2)$ be well defined  (i.e.~obtained  from $A_{\; [b]}(0,0)$ along the two paths reaching $(L_1,L_2)$ from $(0,0)$). We begin with the  cocycle conditions for the  transition functions for $SU(N-k)$ and $SU(k)$
 \begin{eqnarray}
 \Omega_{1 \; [N-k]}(L_2) \; \Omega_{2 \; [N-k]}(0) &=& \left[\begin{array}{cc} e^{i {2 \pi \over N-k} p_{N-k}} I_{N-k} &0 \cr 0 & I_k \end{array}\right]  \; \Omega_{2 \; [N-k]}(L_1) \; \Omega_{1 \; [N-k]}(0) \label{sunktransition} \\
  \Omega_{1 \; [k]}(L_2) \; \Omega_{2 \; [k]}(0) &=&  \left[\begin{array}{cc}  I_{N-k} &0 \cr 0 & e^{i {2 \pi \over k} p_{k}}I_k \end{array}\right]   \; \Omega_{2 \; [k]}(L_1) \; \Omega_{1 \; [k]}(0) \label{suktransition}~.
 \end{eqnarray}
 In $SU(N-k)$ and $SU(k)$ gauge theories, taken in isolation,\footnote{Our theory arises  from the adjoint-Higgs breaking $SU(N)\rightarrow U(1) \times SU(N-k) \times SU(k)$.  The gauge group of the $k$-wall worldvolume theory is  $SU(N-k)\times SU(k) \times U(1) \over \Z_{N-k} \times \Z_k$.} the center elements appearing on the r.h.s. of (\ref{sunktransition}, \ref{suktransition})  equal unity, i.e. $p_{N-k} = p_{k} = 0$. 
 
We also recall that in $SU(N)$ gauge theories, there is a global one-form $\Z_{N}^{(1)}$ center symmetry acting on the transition functions. The boundary conditions for all fields are invariant under this symmetry if there are no fields in the fundamental representation (the boundary conditions for  fundamental fields explicitly break the symmetry).
The same holds for $SU(N-k)$ or $SU(k)$ gauge theories with only adjoint fields.
For example, for $SU(N-k)$ there is a 1-form $\Z_{N-k}^{(1)}$ acting on the $SU(N-k)$ transition functions as
 \begin{eqnarray}
 \label{centerznk}
 \Omega_{i \; [N-k]}  &\rightarrow& z_{i \; [N-k]}\; \Omega_{i \; [N-k]}~, ~i=1,2,
\end{eqnarray}
 where $z_{i \; [N-k]} \in \Z_{N-k}$ is an $x$-independent  constant. The cocycle conditions 
 (\ref{sunktransition}) and all boundary conditions (in a theory without fundamental fields) are invariant under (\ref{centerznk}). The fundamental representation Wilson line operator (\ref{wilsonloop}) is the only one transforming under (\ref{centerznk}). This follows from the explicit appearance of the transition function in its gauge invariant definition, implying $W_{[N-k]} (x_2) \rightarrow z_{1\; [N-k]} W_{[N-k]} (x_2)$; similarly, a loop winding in the $x^2$ direction transforms by $z_{2 \; [N-k]}$. We shall return to the one-form symmetries below when we consider the boundary conditions for the  fermions in the $k$-wall theory. 
 
We next write the $U(1)$ cocycle condition  using the group element $e^{i \omega_i}$ as the $i$-th $U(1)$ transition function. Notice that (\ref{boundary}), written for the $U(1)$ field $A^{N-k}$ as defined in (\ref{kwalldecompositions})  is simply
 \begin{eqnarray}
 \label{boundaryu1}
 A^{N-k} (L_1,x_2) &=& e^{i \omega_1 (x_2)} \left(A^{N-k} (0,x_2) + i d \right)e^{ - i \omega_1 (x_2)}  ~, \nonumber \\
 A^{N-k}(x_1,L_2) &=& e^{i \omega_2 (x_1)} \left(A^{N-k} (x_1,0) + i d \right) e^{ - i \omega_2 (x_1)}  ~,\\
 e^{i \omega_1 (L_2)} \; e^{i \omega_2(0)} &=& e^{i 2 \pi \alpha} \; e^{i \omega_2 (L_1)} \; e^{i \omega_1(0)} \nonumber 
 \end{eqnarray} 
  while on the last line we wrote the $U(1)$  cocycle condition on the torus determined by a  for now arbitrary phase.   
  In terms of the $SU(N)$-embedded representation (\ref{u1transitionfunctions}), the above $U(1)$ cocycle condition becomes
  \begin{eqnarray}
  \label{u1transition}
  \Omega_{1 \; [1]}(L_2) \;  \Omega_{2 \; [1]}(0) =  \left[\begin{array}{cc} e^{i2\pi \alpha \frac{k}{N} \gamma^{(N-k)}} I_{N-k} &0 \\0&  e^{i2\pi \alpha \frac{k -N}{N} \gamma^{(N-k)}} I_{k}  \end{array}\right] \Omega_{2 \; [1]}(L_1) \;  \Omega_{1 \; [1]}(0)~.
  \end{eqnarray}
  
  The main point we want to stress now is that when the $U(1)\times SU(N-k) \times SU(k)$ theory arises from the adjoint-Higgs mechanism from the $SU(N)$ theory, the individual twists (nontrivial phases in the cocycle conditions) in (\ref{sunktransition}, \ref{suktransition}, \ref{u1transition}) do not have to vanish, but can instead conspire to lead to zero twist in $SU(N)$. 
   In other words, we consider the transition functions embedded  in the $SU(N)$ theory and impose the cocycle condition for   vanishing $SU(N)$ twist 
   \begin{eqnarray}
 \label{sunembeddedccocycle}
e^{i {2 \pi \over N} p_N } I_{N}&=&\left[\begin{array}{cc} e^{i \alpha \frac{2 \pi k}{N} \gamma^{(N-k)}} I_{N-k} &0 \\0& e^{i  \alpha \frac{2 \pi (k-N)}{N} \gamma^{(N-k)}} I_{k}   \end{array}\right]\left[\begin{array}{cc} e^{i \frac{2\pi }{N-k}p_{N-k}}I_{N-k} &0\\0 & I_{k}\end{array} \right]\left[\begin{array}{cc} I_{N-k} &0 \\ 0& e^{i \frac{2 \pi}{k} p_k}I_{k} \end{array} \right] \nonumber \\
&=&\left[\begin{array}{cc}  e^{i \alpha \frac{2 \pi k}{N} \gamma^{(N-k)}}e^{i \frac{2\pi }{N-k}p_{N-k}} I_{N-k} &0 \\0 & e^{i  \alpha \frac{2 \pi (k-N)}{N}\gamma^{(N-k)}} e^{i \frac{2 \pi}{k} p_k} I_{k} \end{array} \right]\,.
   \end{eqnarray}
Above, we allowed for  nonvanishing $SU(N)$ twist, $p_N\ne 0$ for future use; imposing vanishing $SU(N)$ twist corresponds to taking $p_N = 0$. 
The  above cocycle conditions, keeping $p_N\ne 0$, have the following general solution for $\alpha$, given below in one of many possible forms 
\begin{eqnarray}
\label{solution34}
\alpha \; \gamma^{(N-k)} \; k (N-k) &=&   p_N (N-k) - N p_{N-k} - N(N-k) m_4 \\
p_{k} &=& p_N - p_{N-k} - k m_5 - (N-k) m_4 ~~ ({\rm mod} \; N)
\end{eqnarray}
where $m_4$ and $m_5$ are arbitrary integers. 
On the top line we chose to express the $U(1)$ twist $\alpha$ in terms of the $SU(N)$ twist $p_N$ and the $SU(N-k)$ twist $p_{N-k}$.  Note that the $SU(k)$ flux is determined by $p_N, p_{N-k}$ and by the integers $m_4$ and $m_5$. 

For further use, we  also write the solution (\ref{solution34}) in an   equivalent way:
\begin{eqnarray} 
\alpha \gamma^{(N-k)} k (N-k) &=& (N-k) p_k - k p_{N-k} - m_4 + m_5 \label{sol1}\\
p_k + p_{N-k} &=&   p_N  - k m_5 - (N-k) m_4 ~~ ({\rm mod} \; N)\label{sol2}  
\end{eqnarray} 

The form (\ref{solution34}) above is convenient to address   our first task:  what $U(1)$ fluxes are possible in the $SU(N)$ theory, i.e. for $p_N=0$ and varying $p_{N-k}$?  The relation between $\alpha$ and $U(1)$ flux follows from using (\ref{boundaryu1}) to express the $U(1)$ flux through the torus in terms of the transition functions
\begin{eqnarray}\label{u1fluxsuN}
\oint {F_{12}^{N-k} dx^1 dx^2 \over 2\pi} = {1 \over 2 \pi}[ \omega_1(L_2)-\omega_1(0) - \omega_2(L_1) + \omega_2(0)] = \alpha~.
\end{eqnarray} where we used the $U(1)$ cocycle condition in the form (\ref{boundaryu1}).
The above equation gives the $U(1)$ flux quantization condition: explicitly, we have, putting $p_N=0$ in (\ref{solution34}):
\begin{equation}
\label{alpha}
\alpha = - \gamma^{(N-k)} (p_{N-k} + (N-k) m_4)~, 
\end{equation}
i.e.~completely equivalent to eqn.~(\ref{solution2}) from the main text, which was used to 
conclude that the $k$-wall theory has a $\Z_{2N}$ anomaly free global symmetry inherited from  the bulk theory. Another derivation of flux 	quantization, using constant flux backgrounds, is given in the main text, Section \ref{section:constantflux}. 

Before continuing our discussion of 't Hooft flux backgrounds, we recall that the $2$-D fermions $\psi_\pm$ also  satisfy a consistency condition on the corners of the torus (single valuedness of $\psi_\pm(L_1,L_2)$)  given by \begin{equation}\label{fermioncocycle}
\psi_\pm (L_1,L_2) = \Omega_1(L_2) \; \Omega_2(0) \circ \psi_\pm(0,0) =   \Omega_2(L_1) \;\Omega_1(0) \circ \psi_\pm(0,0),\end{equation} where $\circ$ denotes the action of the $U(1) \times SU(N-k) \times SU(k)$ gauge transformations on the fermions and $\Omega$ collectively denotes the transition functions for the three gauge groups. Recalling the discussion after (\ref{lambdabetaeqn},\ref{lambdaminusbetaeqn}) of the gauge transformation properties of the fermions, we have explicitly  $\Omega \circ \psi_+$ $=$ $e^{i \omega \gamma^{(N-k)}} \; \Omega_{[N-k]} \; \psi_+ \; \Omega_{[k]}^\dagger$,  $\Omega \circ \psi_-$ $=$ $\e^{- i \omega \gamma^{(N-k)}} \; \Omega_{[k]} \; \psi_+ \;\Omega_{[N-k]}^\dagger$. Thus the cocycle condition  for $\psi_+$  (the $\psi_-$ condition does not bring new constraints) becomes  
\begin{eqnarray}\label{fermioncocycle2}
\psi_+(L_1,L_2)& & \\
=&&e^{i \omega_1(L_2) \gamma^{(N-k)}}\;  e^{i \omega_2(0) \gamma^{(N-k)}} \; \Omega_{1\; [N-k]}(L_2)  \; \Omega_{2\; [N-k]}(0) \; \psi_+(0,0) \; \Omega_{2\; [k]}^\dagger(0)  \; \Omega_{1\; [k]}^\dagger(L_2) \nonumber    \\ 
=&&e^{i \omega_2(L_1) \gamma^{(N-k)}}\;  e^{i \omega_1(0) \gamma^{(N-k)}} \; \Omega_{2\; [N-k]}(L_1) \;  \Omega_{1\; [N-k]}(0) \; \psi_+(0,0) \; \Omega_{1\; [k]}^\dagger(0)  \; \Omega_{2\; [k]}^\dagger(L_1)~. \nonumber\end{eqnarray}

 The bulk $SU(N)$ theory has a 
 $\Z_{N}^{(1)}$ 1-form global symmetry. The part of this 1-form symmetry projected to the $2$-D $k$-wall worldvolume  is unbroken by the adjoint Higgsing and should still be manifest as a symmetry acting on  transition functions, as described earlier. The transition functions  $\Omega_{i \; [1]}, \Omega_{i \; [N-k]}, \Omega_{i \; [k]}$, collectively denoted by $\Omega_i$,   embedded into $SU(N)$  are:
\begin{eqnarray}\label{sunomegai}
\Omega_i = 
 \left[\begin{array}{cc} e^{i2\pi \omega_i \frac{k}{N} \gamma^{(N-k)}} \Omega_{i\; [N-k]} &0 \\0&  e^{i2\pi \omega_i \frac{k -N}{N} \gamma^{(N-k)}}  \Omega_{i\; [k]}  \end{array}\right] ~,~ i=1,2.
 \end{eqnarray}
Consider now the  1-form symmetry action (say, in the $x^i$ direction)  on the transition functions:
\begin{eqnarray}\label{znoneformaction}
 e^{i \omega_{i}} &\rightarrow& e^{i 2 \pi \zeta_{(i)}}\; e^{i \omega_i} \nonumber  , \\
   \Omega_{i\; [N-k]} &\rightarrow& e^{i {2 \pi \over N-k} q_{(i)  N-k}} \;\Omega_{i\; [N-k]} ,\\
  \Omega_{i\; [k]} &\rightarrow& e^{i {2 \pi \over  k} q_{(i)  k}}\; \Omega_{i\; [ k]} ~\nonumber .
  \end{eqnarray} In order that  (\ref{sunomegai}) transform by a $\Z_N$ phase $z_{i [N]} \equiv e^{i {2 \pi \over N} p_{(i)}}$, i.e., $\Omega_i \rightarrow e^{i {2 \pi \over N} p_{(i)}} \Omega_i$ as appropriate to a 1-form $\Z_N$ symmetry,  the following conditions should hold\footnote{Note also that that  (\ref{zn}) with arbitrary $p_{(i)}$ (not necessarily quantized) represents the most general one-form  transformations (\ref{znoneformaction}) that leave the fermion cocycle conditions (\ref{fermioncocycle2}) invariant. This follows upon inspection; also recall that the cocycle conditions have to respect the $1$-form symmetry.  Hence, this is the most general ansatz for an unbroken $1$-form symmetry of the $k$-wall theory. Using the solution given in  (\ref{solution34}), also valid for arbitrary $p_{(i)}$, one can show that the $\Z_N^{(1)}$ found below is the only $1$-form symmetry in the $k$ wall theory.}
\begin{eqnarray}\label{zn}
e^{i 2 \pi \zeta_{(i)} {k \over N} \gamma^{(N-k)}}\; e^{i {2 \pi \over N-k} q_{(i)   N-k}}  &=& e^{ i {2 \pi \over N}~ p_{(i)}} \\
e^{i 2 \pi \zeta_{(i)} {k-N \over N}\gamma^{(N-k)}} \; e^{i {2 \pi \over k} q_{(i)   k}}  &=& e^{ i {2 \pi \over N}~ p_{(i)}} \nonumber \end{eqnarray}
These conditions are formally equivalent to the ones in (\ref{sunembeddedccocycle}) determining the cocycle conditions for the $U(1)$ field (we stress that here they have a different meaning), upon the replacement $\alpha \rightarrow \zeta_{(i)}$, $p_{N-k} \rightarrow q_{(i) \; N-k}$, $p_k \rightarrow q_{(i) \; k}$, $p_N \rightarrow p_{(i)}$, and we can borrow their solutions. We express the solution (\ref{sol1}, \ref{sol2}) for $\zeta_{(i)}$ as follows (setting $m_4$ and $m_5$ to zero)
\begin{eqnarray} \label{zn2}
  \zeta_{(i)}   \gamma^{(N-k)}   &=& {q_{(i) k} \over k} - {q_{(i) N-k} \over N-k}   \label{a1} \\
q_{(i) k} + q_{(i) N-k} &=&   p_{(i)}   ~~ ({\rm mod} \; N)\label{a32}  ~\nonumber .
\end{eqnarray}
These conditions ensure that (\ref{zn}) represent the action of a  $\Z_N$ symmetry. It is easy to see that the $k$-wall fermion boundary  conditions (\ref{fermioncocycle2}),   are invariant under the $\Z_N^{(1)}$ symmetry action (\ref{zn}, \ref{zn2}). 
To further check (\ref{zn}, \ref{zn2}), consider the action on a fundamental $SU(N)$ Wilson loop. 
 A fundamental of $SU(N)$ decomposes  into two representations  under the unbroken $ U(1)\times SU(N-k) \times SU(k)$ gauge group: $q_1 \sim (, {k \over N} \gamma^{(N-k)}, \Box, {\bf 1})$ and $q_2\sim (  ({k -N \over N}) \gamma^{(N-k)}, {\bf 1}, \Box)$, as seen from (\ref{tildeH}, \ref{kwallgenerators}). A Wilson loop of, say $q_1$, along the $x^1$ direction of the $k$-wall wordlvolume has the form 
\begin{equation}\label{wilsonloop1}
W_{q_1} (x_2) = \tr( e^{i \int\limits_{ 0}^{L_1} A_{1 \; [N-k]} (x_1,x_2) dx^1} \Omega_{1 \; [N-k]} (x_2) ) \; e^{i {k \over N} \gamma^{(N-k)}  \int\limits_{ 0}^{L_1} A_1^{N-k} (x_1,x_2) dx^1}e^{i \omega_{1} (x_2) 
 {k \over N} \gamma^{(N-k)}} \end{equation}
where we recalled (\ref{wilsonloop})  and inserted appropriate $SU(N-k)$ and $U(1)$ transition functions $\Omega_{1 \; [N-k]}, e^{i \omega_1 {k \over N} \gamma^{(N-k)}}$. Under (\ref{znoneformaction}) with $i=1$, we have $W_{q_1} (x_2)$ $\rightarrow$ $e^{i 2 \pi \zeta_{(1)} {k \over N} \gamma^{(N-k)}}$ $e^{i {2 \pi \over N-k} q_{(1)   N-k}}   W_{q_1} (x_2)$ $= $  $e^{i {2 \pi \over N} p_{(1)  }}   W_{q_1} (x_2)$,  using (\ref{zn}, \ref{zn2}), as appropriate for a $\Z_N^{(1)}$ 1-form symmetry.

  \section{'t Hooft fluxes and projection of constant flux backgrounds}
  \label{appendix:thooftprojection}
  
  The discussion that follows is an explicit demonstration that the constant flux backgrounds (\ref{fluxes1}) with $\bm u = \bm \omega^b$ are   examples of backgrounds obeying the twisted boundary conditions (\ref{sunembeddedccocycle}) with nonzero $SU(N)$ twists, $p_N \ne 0$. 
  
  We begin with (\ref{fluxes3}), which gives the $SU(N)$ twist associated to the constant flux background  (\ref{fluxes1}) as 
  \begin{equation} \label{twist1}
  t(\bm u) = e^{i 2 \pi \bm u \cdot \bm H} . 
 \end{equation}
  Here, we project that twist $t$ onto  $SU(N-k)$ and $SU(k)$. The projection on the $U(1)$ factor was done before, see Section \ref{section:constantflux}. The result is:
    \begin{equation} \label{twist10}
  t(\bm u)_{U(1)} =  (e^{ - i {2 \pi \over N-k} \sum\limits_{A=N-k+1}^N \bm u \cdot \bm \nu^A} I_{N-k}, \; e^{   i {2 \pi \over k} \sum\limits_{A=N-k+1}^N \bm u \cdot \bm \nu^A} I_k). 
 \end{equation}
 In particular, for $\bm u =  \bm \alpha^b$, there we found that $\sum\limits_{A=N-k+1}^N \bm \alpha^b \cdot \bm \nu^A = - \delta^{b, N-k}$, thus
   \begin{equation}
    \label{twist101}
  t(\bm \alpha^{N-k})_{U(1)} = (e^{  i {2 \pi \over N-k} } I_{N-k}, \; e^{  - i {2 \pi \over k}  } I_k),~~~  t(\bm \alpha^{b \ne N-k})_{U(1)} = I_N.
 \end{equation}
 
      To continue, we introduce some notation. We let indices $\tilde{a}, \tilde{b},... = 1, \ldots N-k-1$, and $\tilde{A}, \tilde{B},... = 1, \ldots N-k$. Primed indices are  $a', b'... = 1, \ldots k-1$, and $A', B',... = 1, \ldots k$. Finally, as before, 
 $a, b,...= 1,\ldots N-1$ and $A,B,...= 1, \ldots N$.
We denote the $SU(N-k)$ Cartan generators by $\tilde{  H}^{\tilde{a}}$. These are $(N-k)\times(N-k) $ diagonal matrices whose $\tilde{A}$-th eigenvalue we denote by  
 $(\tilde{  H}^{\tilde{a}})^{\tilde{A}} = \lambda^{\tilde{a} \tilde{A}} = (\bm \nu^{\tilde{A}})^{\tilde{a}}$.
The $SU(k)$ generators are ${  H'}^{a'}$, $k \times k$ matrices with eigenvalues $({  H'}^{a'})^{A'}= (\bm \nu^{{A'}})^{{a'}}$. 
The set of $SU(N)$ generators consisting of $\tilde{H}^{\tilde{a}}$ , ${H'}^{a'}$, $\tilde{H}^{N-k}$,   embedded into $SU(N)$ as in (\ref{kwallgenerators}),  are an orthonormal basis of $SU(N)$ generators suited for the $SU(N) \rightarrow U(1) \times SU(N-k) \times SU(k)$ breaking pattern. 
 
 We first project (\ref{twist1}) onto the $SU(N-k)$ subgroup. We define the projection onto $SU(N-k)$ as\footnote{Taking the trace entails embedding the generators into $SU(N)$ as in (\ref{kwallgenerators}). Expressing the result through the weights  yields $\tr (H^a \tilde{H}^{\tilde{a}}) = 
 \sum\limits_{\tilde B=1}^{N-k} (\bm \nu^{\tilde B})^{a} (\bm \nu^{\tilde B})^{\tilde a} $.}
 \begin{equation}
 \label{twist2}
 t(\bm u)_{SU(N-k)} \equiv e^{i 2 \pi   u^a \tr (H^a \tilde{H}^{\tilde{a}}) H^{\tilde{a}}}~,
 \end{equation}
 hence the $\tilde{A}$-th eigenvalue of $t_{N-k}$ is 
  \begin{equation}
 \label{twist21}
 t(\bm u)_{SU(N-k)}^{\tilde{A}} \equiv e^{i 2 \pi   u^a \tr (H^a \tilde{H}^{\tilde{a}}) (H^{\tilde{a}})^{\tilde{A}}}~= e^{i 2\pi u^a (\bm \nu^{\tilde B})^{a} (\bm \nu^{\tilde B})^{\tilde a}   (\bm \nu^{\tilde A})^{\tilde a}}~,
 \end{equation}
 where the sums in the exponent in the last term are over the appropriate ranges indicated above ($a=1,...N-1$, $\tilde{B}=1,...,N-k$, $\tilde{a}=1,...N-k-1$). We now note that  
 $(\bm \nu^{\tilde B})^{\tilde a}   (\bm \nu^{\tilde A})^{\tilde a} = \delta^{\tilde{A} \tilde{B}} - {1 \over N-k}$ 
 , as $(\bm \nu^{\tilde A})^{\tilde a}$ are the weights of the fundamental for $SU(N-k)$. Thus, we have
  \begin{equation}
 \label{twist22}
 t(\bm u)_{SU(N-k)}^{\tilde{A}} = e^{i \; 2\pi u^a (\bm \nu^{\tilde A})^{a}    - i \; {2\pi \over N-k}\;u^a \sum\limits_{\tilde{B}=1}^{N-k} (\bm \nu^{\tilde B})^{a}  }~,
 \end{equation}
 where we explicitly indicated the sum over $\tilde B$ in the second term  (the sum over $a$ is still in the appropriate range $1,...,N-1$). 
 
 We now imagine that $\bm u = \bm \alpha^b = \bm \nu^b -\bm \nu^{b+1}$ is a $SU(N)$ root vector. Thus, recalling that $\bm\alpha^b \cdot  \bm \nu^{\tilde A} \in \Z$, 
\begin{eqnarray}
 \label{twist23}
 t(\bm \alpha^b)_{SU(N-k)}^{\tilde{A}} &=& e^{i \; 2\pi (\bm\alpha^b)^a (\bm \nu^{\tilde A})^{a}    - i \; {2\pi \over N-k}\;(\bm\alpha_b)^a \sum\limits_{\tilde{B}=1}^{N-k} (\bm \nu^{\tilde B})^{a}  }~ = e^{    - i \; {2\pi \over N-k}\; \sum\limits_{\tilde{B}=1}^{N-k}(\bm \nu^b - \bm \nu^{b+1}) \cdot \bm \nu^{\tilde B}   }~\nonumber\\
 &=& e^{    - i \; {2\pi \over N-k}\; \sum\limits_{\tilde{B}=1}^{N-k}(\delta^{b \tilde{B}} - \delta^{b+1 \tilde{B}} )} = e^{    - i \; {2\pi \over N-k}\delta^{b N-k}}~.
 \end{eqnarray}
As promised earlier, for $\bm u = \bm \alpha^b$ there is a nontrivial twist in $SU(N-k)$, cancelled by the twist in the $U(1)$ factor, for $b = N-k$ only (recall discussion after (\ref{fluxes6})).

Next, we consider nontrivial 't Hooft fluxes and take $\bm u = \bm \omega^b = \sum\limits_{B=1}^b \bm \nu^B$. The $SU(N-k)$ twist becomes
  \begin{eqnarray}
 \label{twist24}
 t(\bm \omega^b)_{SU(N-k)}^{\tilde{A}} &=& e^{i \; 2\pi  \sum\limits_{B=1}^b \bm \nu^B \cdot \bm \nu^{\tilde A}    - i \; {2\pi \over N-k}\; \sum\limits_{\tilde{B}=1}^{N-k} \sum\limits_{B=1}^b  \bm \nu^B \cdot \bm \nu^{\tilde B}   }~\\
 &=&e^{i \; 2\pi  \sum\limits_{B=1}^b(\delta^{B \tilde A} - {1 \over N})   - i \; {2\pi \over N-k}\; \sum\limits_{\tilde{B}=1}^{N-k} \sum\limits_{B=1}^b   (\delta^{B  \tilde B} - {1 \over N})  } ~ \nonumber \\
 &=&e^{   - i \; {2\pi \over N-k}\; \sum\limits_{\tilde{B}=1}^{N-k} \sum\limits_{B=1}^b \delta^{B  \tilde B}   } = e^{- i {2 \pi \over N-k} {\rm min}(b, N-k)}
 \end{eqnarray}
 Before calculating this expression, we notice that this is a twist in the center of $SU(N-k)$, for all $\bm \omega^b$ (as it is $\tilde{A}$-independent) and that it is only non-unity for $b < N-k$.
 
 Let us now calculate the same twist for the $U(1)$ factor (\ref{twist10}). We need 
 \begin{eqnarray}
 \label{twist03}
 \sum\limits_{A = N-k+1}^N \sum\limits_{B=1}^b \bm \nu^B \cdot \bm \nu^A  =\sum\limits_{A = N-k+1}^N \sum\limits_{B=1}^b  (\delta^{AB} - {1 \over N})=   \left\{ \begin{array}{cc} - {k b \over N}  &, b \le N-k \cr (N-k)({b \over N} - 1)  &, b > N-k \end{array}. \right.
 \end{eqnarray}
 Thus we find
   \begin{equation} \label{twist04}
  t(\bm \omega^b)_{U(1)} =    \left\{ \begin{array}{ccc} ( e^{  i {2 \pi \over N-k } {k \over N} b} \; I_{N-k},& \; e^{- i {2 \pi b \over N}} \; I_k )   &, b \le N-k \cr   ( e^{ - i {2 \pi \over N }  b} \; I_{N-k},& \; e^{  i {2 \pi   \over k} {N-k \over N} (b - N)}\; I_k )  &, b > N-k \end{array}. \right.
  \end{equation}
  We notice that the $SU(N-k)$ twist in (\ref{twist24}) and the $U(1)$ twist above conspire to produce a  twist, $e^{- i {2 \pi b \over N}}$,  in the center of $SU(N)$  for all values of $b$.   
  
  Finally, we show that the same ``conspiracy" holds for the $SU(k)$ and $U(1)$ twists. The projection on $SU(k)$ is  
  \begin{equation}
 \label{twist5}
 t(\bm u)_{SU(k)} \equiv e^{i 2 \pi   u^a \tr (H^a  {H}^{' a'}) H^{a'}}~,
 \end{equation}
 whose $A'$-th (recall $A', B'... = 1,\ldots k$, $a', b'... =1,\ldots k-1$) component (recalling  $\bm\nu^{A'})^{a'}$ are the $SU(k)$ weights of the fundamental) is
 \begin{eqnarray}
 \label{twist51}
 t(\bm u)_{SU(k)}^{A'} &=& e^{i 2 \pi   (\bm u  \cdot \bm\nu^{N-k+B'}) (\bm\nu^{B'})^{a'} (\bm\nu^{A'})^{a'}} = e^{i 2 \pi  \sum\limits_{B'=1}^k (\bm u  \cdot \bm\nu^{N-k+B'}) (\delta^{A' B'} - {1 \over k})} \nonumber \\
 &=&e^{i 2 \pi  \bm u  \cdot \bm\nu^{N-k+A'}   - i {2 \pi \over k}  \sum\limits_{B'=1}^k  \bm u  \cdot \bm\nu^{N-k+B'}   }
 \end{eqnarray}
We first take a root lattice $\bm u = \bm \alpha^b = \bm \nu^b -\bm \nu^{b+1}$ and find 
\begin{eqnarray}
 \label{twist52}
 t(\bm \alpha^b)_{SU(k)}^{A'} 
 &=&e^{i 2 \pi  (\bm \nu^b -\bm \nu^{b+1})  \cdot \bm\nu^{N-k+A'}   - i {2 \pi \over k}  \sum\limits_{B'=1}^k  (\bm \nu^b -\bm \nu^{b+1})   \cdot \bm\nu^{N-k+B'}   }\nonumber  \\
  &=&e^{  - i {2 \pi \over k}  \sum\limits_{A=N-k+1}^N   (\delta^{b, A} - \delta^{b+1, A})} = e^{   i {2 \pi \over k} \delta^{b, N-k}}~.
 \end{eqnarray}
Thus, as expected, for root lattice $\bm u$, the $SU(k)$ twist exactly cancels the $\sim I_{k}$ part of the $U(1)$ twist (\ref{twist101}). We finally   take a  weight lattice $\bm u$ and find from (\ref{twist51})
\begin{eqnarray}
 \label{twist53}
 t(\bm \omega^b)_{SU(k)}^{A'}  
 &=&e^{i 2 \pi     \sum\limits_{p=1}^b \bm \nu^p  \cdot \bm\nu^{N-k+A'}  - i {2 \pi \over k}  \sum\limits_{B'=1}^k  \sum\limits_{p=1}^b \bm \nu^p  \cdot \bm\nu^{N-k+B'}   } \nonumber \\
 &=&e^{i 2 \pi    \sum\limits_{p=1}^b (\delta^{p, N-k+A'} - {1 \over N})- i {2 \pi \over k}  \sum\limits_{B'=1}^k  \sum\limits_{p=1}^b (\delta^{p, N-k+B'} - {1 \over N})    } \nonumber \\
  &=&e^{   - i {2 \pi \over k}  \sum\limits_{A=N-k+1}^N  \sum\limits_{p=1}^b  \delta^{p,A}      } =    \left\{\begin{array}{c} 1, b \le N-k \cr e^{ - i {2 \pi \over k} (b -  N)}, b > N-k\end{array} \right. ,	   
 \end{eqnarray}
thus, comparing with the $U(1)$ twist (\ref{twist04}) we see that the $U(1)$ twist cancels the $\Z_k$ $SU(k)$ twist and a $\Z_N$ twist $e^{- i {2 \pi b \over N}}$ is left over for all values of $b$. 

\bibliography{DWsuNbiblio.bib}
\bibliographystyle{JHEP}



\end{document}